\documentclass[%
 aip,
 amsmath,amssymb,
preprint,%
]{revtex4-1}

\usepackage{graphicx}
\usepackage{dcolumn}
\usepackage{bm}

\usepackage[utf8]{inputenc}
\usepackage[T1]{fontenc}
\usepackage{mathptmx}
\usepackage{etoolbox}
\usepackage{xcolor}
\usepackage{tikz}
\usepackage{multirow}
\usepackage[caption=false]{subfig}
\usepackage{hyperref}
\hypersetup{hidelinks}
\usepackage[capitalise]{cleveref}
\usepackage[normalem]{ulem} 
\usepackage{CJK} 
\usepackage{xspace}
\usepackage{framed}

\usepackage{glossaries-extra}
\setabbreviationstyle[acronym]{long-short}
\glssetcategoryattribute{acronym}{nohyperfirst}{true}
\newacronym{RANS}{RANS}{Reynolds-averaged Navier-Stokes}
\newacronym{CFD}{CFD}{Computational Fluid Dynamics}
\newacronym{EPF}{EPF}{Eigenspace Perturbation Framework}
\newacronym{AIM}{AIM}{Anisotropy Invariant Map}
\newacronym{ABM}{ABM}{Anisotropy Barycentric Map}
\newacronym{DLR}{DLR}{German Aerospace Center}
\newacronym{UQ}{UQ}{Uncertainty Quantification}
\newacronym{V&V}{V\&V}{Verification and Validation}
\newacronym{LEVM}{LEVM}{Linear Eddy Viscosity Models}
\newacronym{PCS}{PCS}{Principal Coordinate System}
\newacronym{DNS}{DNS}{Direct Numerical Simulation}
\newacronym{QoI}{QoI}{Quantities of Interest}

\def\CorrectionsReviewOne{2}     

\ifnum\CorrectionsReviewOne=1    
    \newcommand{\cancelText}[1]{{\textcolor{red}{\sout{#1}}}}%
    \newcommand{\newText}[1]{{\textcolor{blue}{#1}}}%
\fi

\makeatletter
\ifnum\CorrectionsReviewOne=2    
    \newcommand{\cancelText}[1]{\@bsphack\@esphack}%
    \newcommand{\newText}[1]{#1}%
\fi

\def\CorrectionsReviewTwo{2}     

\ifnum\CorrectionsReviewTwo=1    
    \newcommand{\cancelTextTwo}[1]{{\textcolor{red}{\sout{#1}}}}%
    \newcommand{\newTextTwo}[1]{{\textcolor{blue}{#1}}}%
\fi

\makeatletter
\ifnum\CorrectionsReviewTwo=2    
    \newcommand{\cancelTextTwo}[1]{\@bsphack\@esphack}%
    \newcommand{\newTextTwo}[1]{#1}%
\fi

\makeatletter
\def\@email#1#2{%
 \endgroup
 \patchcmd{\titleblock@produce}
  {\frontmatter@RRAPformat}
  {\frontmatter@RRAPformat{\produce@RRAP{*#1\href{mailto:#2}{#2}}}\frontmatter@RRAPformat}
  {}{}
}%

\makeatother
\begin{document}

\preprint{AIP/Physics of Fluids}

\title[Improved self-consistency of the Reynolds stress tensor eigenspace perturbation for UQ]{Improved self-consistency of the Reynolds stress tensor eigenspace perturbation for Uncertainty Quantification}
\author{Marcel Matha*}
\author{Christian Morsbach}%
 \email{marcel.matha@dlr.de.}
\affiliation{ 
\gls{DLR}, Linder Höhe, 51147 Cologne, Germany}


\date{\today}

\begin{abstract}
The limitations of turbulence closure models in the context of Reynolds-averaged Navier-Stokes (RANS) simulations play a significant part in contributing to the uncertainty of Computational Fluid Dynamics (CFD). 
Perturbing the spectral representation of the Reynolds stress tensor within physical limits is common practice in several commercial and open-source CFD solvers, in order to obtain estimates for the epistemic uncertainties of RANS turbulence models.\cancelText{We point out that the need for moderating the perturbation due to upcoming stability issues of the solver in the common implementation leads} \newText{Recent research revealed, that there is a need for moderating the amount of perturbed Reynolds stress tensor tensor to be considered due to upcoming stability issues of the solver. In this paper we point out that the consequent common implementation can lead} to unintended states of the resulting perturbed Reynolds stress tensor. The combination of eigenvector perturbation and moderation factor may actually result in moderated eigenvalues, which are not linearly dependent on the originally unperturbed and fully perturbed eigenvalues anymore. Hence, the computational implementation is no longer in accordance with the conceptual idea of the Eigenspace Perturbation Framework. \cancelText{In this paper w}\newText{W}e verify the implementation of the conceptual description with respect to its self-consistency. Adequately representing the basic concept results in formulating a computational implementation to improve self-consistency of the Reynolds stress tensor perturbation.

\end{abstract}

\maketitle

\section{\label{sec:introduction} Introduction \protect}

Industrial aerodynamic designs increasingly rely on numerical analysis based on flow simulations using \gls{CFD} software. Such industrial applications usually feature turbulent flows. Due to its cost- and time-effective solution procedure, Reynolds-averaged Navier-Stokes (RANS) equations are an appropriate approach for design optimizations and virtual certification. Unfortunately, the \gls{RANS} equations are not closed and, hence, require the determination of the second-moment Reynolds stress tensor. In this context, the Reynolds stress tensor is approximated using turbulence models. These models make assumptions regarding the relationship between the Reynolds stresses and available mean flow quantities, such as the mean velocity gradients, which limit their applicability in terms of accuracy on the one hand. On the other hand, the assumptions made\cancelText{ during} \newText{in the} formulation of closure models inevitably lead to uncertainties as soon as their range of validity is left\cancelText{ and their}\newText{. The} quantification \newText{of these model-form uncertainties} for industrial purposes is a demanding task in general.\\ 
Several approaches seek to account for these uncertainties at different modeling levels~\cite{Duraisamy, XIAO20191}.
We focus on the \gls{EPF}~\cite{Emory, Iaccarino}, which estimates the predictive uncertainty due to limitations in the turbulence model structure, namely its epistemic uncertainty. The \gls{EPF} is purely physics-based and introduces a series of perturbations to the shape, alignment and size of the modeled Reynolds stress ellipsoid to estimate its uncertainty. Because of its straight forward implementation, the \gls{EPF} has been used in diverse areas of application such as mechanical engineering~\cite{razaaly2019optimization}, aerospace engineering~\cite{mishra2017uncertainty, cook2019optimization, mishra2020design, Chu2022a, Chu2022}, civil engineering~\cite{garcia2014quantifying, Lamberti}, wind farm design~\cite{EidiDataFree, Hornshoj}, etc. 
  The \gls{EPF} is the foundation of recent confidence-based design under uncertainty approaches~\cite{gori2022confidence}. There have been studies showing the potential to optimize it using data driven machine learning approaches~\cite{Heyse, Eidi2022} and it has been applied for the virtual certification of aircraft designs~\cite{Mukhopadhaya,nigam2021toolset}.
The \gls{EPF} has been integrated into several open and closed source flow solvers~\cite{Edeling, Gorle2019, MishraSU2,MathaCF}. This range of applications emphasizes the importance of the \gls{EPF}. Imperfections in the \gls{EPF} can have a cascading ramification to all these applications and fields.
\\
There is need for \gls{V&V} for such novel methodologies. Validation focuses on the agreement of the computational simulation with physical reality~\cite{ oberkampf2002verification}, which has been done for the \gls{EPF} in the aforementioned studies. On the other hand, verification focuses on the correctness of the programming and computational implementation of the conceptual model~\cite{ stern2001comprehensive}. For the \gls{EPF}, this verification would involve the theory behind the conceptual model and the computational implementation. The theoretical foundations of the Reynolds stress tensor perturbations have been analyzed in detail~\cite{mishra2019theoretical}. In this investigation, we focus on the computational implementation of the \gls{EPF}, analyzing the consistency between the envisioned conceptual model and the actually implemented computational model.\\
In order to estimate the epistemic uncertainty for future design applications with respect to turbulence closure model, we review the current implementation of the framework in \newText{\gls{DLR}'s \gls{CFD} solver suite }\textit{TRACE}\cancelText{, which is in accordance with other flow solvers}. Especially, we focus on the motivation, implementation and effects of applying a moderation factor $f$, which serves to mitigate the amount of perturbation and aid numerical convergence of \gls{CFD} solution~\cite{MishraSU2, MathaCF} \newText{(in some publications $f$ is called under-relaxation factor)}. The present investigation reveals a shortcoming when combining the eigenspace perturbation of the Reynolds stress tensor with the moderation factor, which has not yet been addressed in literature. On this basis, we formulate a way of improving self-consistency of the \gls{EPF} and recovering its originally intended, physically meaningful idea in the present paper. Such self-consistency adherence is an essential component of the verification assessment stage of \gls{V&V}~~\cite{roache1998verification} in order to ensure agreement between the conceptual\cancelText{ model} and the computational model (numerical implementation), thus ensuring verification as outlined by AIAA\cancelText{ G-077-1998} \newText{CFD Committee}~\cite{computational1998guide}.\\
The paper is structured as follows: \cref{sec:Method} introduces the Reynolds stress tensor's eigenspace perturbation\cancelText{ in general}. 
We\cancelText{ provide} \newText{describe} the fundamental motivation, the mathematical background and the deduced practical implementation of the \gls{EPF}. In \cref{sec:eigdecomposition}, we present the conceptual idea to apply an eigenspace decomposition of the anisotropy tensor. On this basis, the evident choice to perturb the eigenvalues and eigenvectors within physical limits is demonstrated from a practical engineering perspective in \cref{sec:perturbation}. Propagating these limiting states of turbulence enables a \gls{CFD} practitioner to estimate the model-form uncertainty for certain \gls{QoI} with respect to the underlying turbulence model.
Finally, we point out an inconsistency in the prevailing computational implementation of the eigenspace perturbation in \gls{CFD} solvers\cancelText{ in \cref{sec:needForModeration} and \cref{sec:unwantedEffect}} and suggest an alternative self-consistent formulation in\cancelText{ in \cref{sec:proposedApproach}} \newText{\cref{sec:self-consistentFomrulation}. The uncertainty estimation for simulations of a turbulent boundary layer serve to demonstrate the envisioned benefits of the proposed consistent implementation of the \gls{EPF} in \cref{sec:channelFlowExample}}.
\cref{sec:conclusion} summarizes the findings of the paper and assesses their significance for future applications. 

\section{\label{sec:Method} Reynolds stress tensor perturbation to estimate uncertainties\protect}

\subsection{\label{sec:eigdecomposition} Reynolds stress anisotropy and visualization}

\begin{figure}[htb]
\includegraphics[width=0.4\textwidth, trim=2.5cm 3.25cm 1.45cm 2.5cm, clip=True]{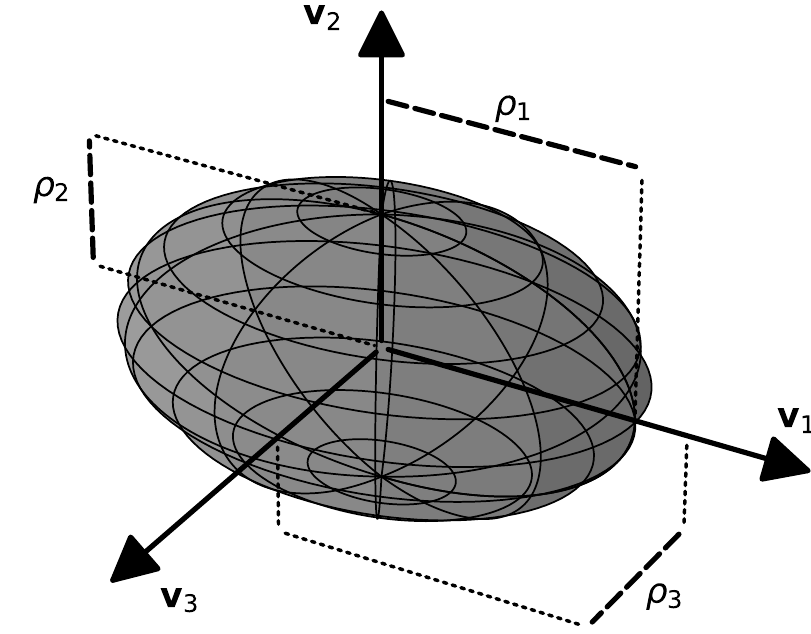}%
\caption{\label{fig:tensorEllipsoid}Representation of tensor as ellipsoid; eigenvalues $\rho_k$ and eigenvectors $v_k$ are highlighted.}
\end{figure}

The symmetric, positive semi-definite Reynolds stress tensor $\tau_{ij} = \overline{u_i' u_j'}$ needs to be determined by turbulence models in order to close the \gls{RANS} equations. It can be decomposed into an anisotropy tensor $a_{ij}$ and an isotropic part 
\begin{equation}
	\label{eq:decompositionTau}
		\tau_{ij} = k \left(a_{ij} + \frac{2}{3}\delta_{ij}\right) \ \text{,}
\end{equation}
where the turbulent kinetic energy is defined as $k = \frac{1}{2} \tau_{kk}$ and summation over recurring indices within a product is implied.
As the Reynolds stress tensor and its symmetric anisotropic part only contain real entries, they are diagonalizable. Thus, based on an eigenspace decomposition, the anisotropy tensor can be expressed as
\begin{equation}
	\label{eq:spectralDecompositionAnisotropy}
		a_{ij} = v_{in} \Lambda_{nl} v_{jl} \ \text{.}
\end{equation} 
The orthonormal eigenvectors form the \gls{PCS} and can be written as a matrix $v_{in}$ while the traceless diagonal matrix $\Lambda_{nl}$ contains the corresponding ordered eigenvalues $\lambda_k$ with respect to $a_{ij}|_\mathrm{PCS}$. Because of the definition of the anisotropy tensor in \cref{eq:decompositionTau}, Reynolds stress and anisotropy tensor share the same eigenvectors while the eigenvalues of the Reynolds stress tensor are $\rho_k = k (\lambda_k + 2/3)$. 
\cancelText{Altogether} \newText{Consequently}, the eigenvalues and the eigenvectors represent the shape and the orientation of the positive semi-definite (3,3)-tensor and can be visualized as an ellipsoid (see \cref{fig:tensorEllipsoid}).

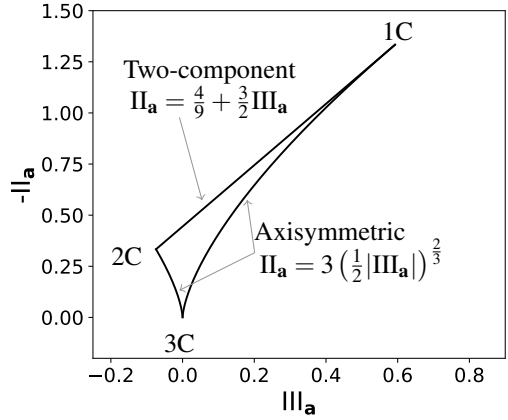
\begin{figure}
\centering
    \begin{tikzpicture}
        \node[] at (0,0) {
                \includegraphics[width=0.45\textwidth, trim=0cm 0.8cm 0cm 0cm, clip=True]{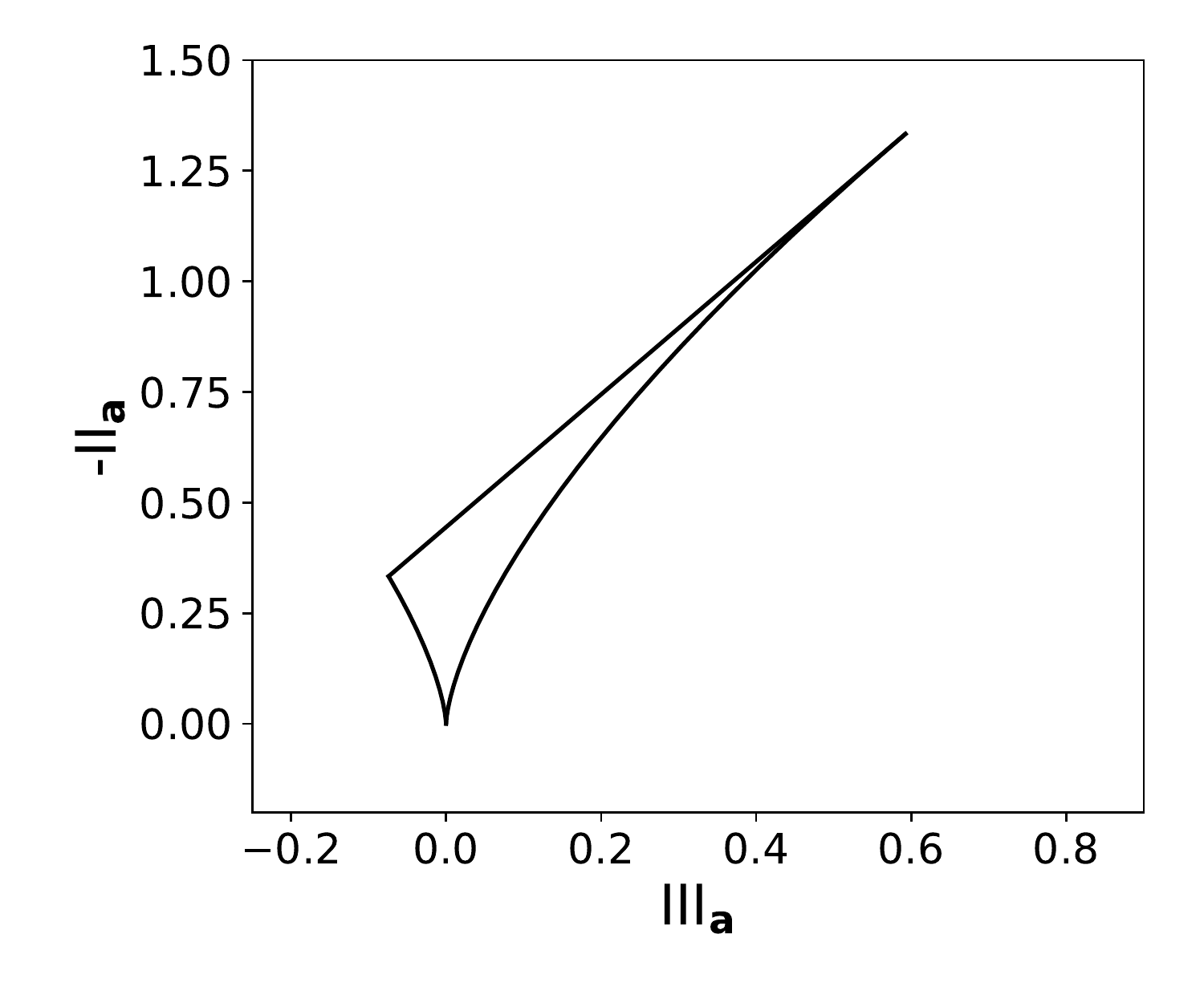}};
        \draw [gray!80!white, <-]( -0.1,0) -- (-0.0,-0.7) node[] {};
        \draw [gray!80!white, <-](-1.0,-1.2) -- ( -0.0,-0.7) node[] {};
        \node[black] at (1, -0.45) {\footnotesize Axisymmetric};
        \node[black] at (1.3, -0.8) {\footnotesize $\text{II}_{\mathbf{a}} = 3\left(\frac{1}{2}|\text{III}_{\mathbf{a}}|\right)^\frac{2}{3}$};
        \draw [gray!80!white, ->]( -1,1.1) -- (-0.7,0.0) node[] {};
        \node[black] at (-0.6, 1.7) {\footnotesize Two-component};
        \node[black] at (-0.6, 1.3) {\footnotesize $\text{II}_{\mathbf{a}} = \frac{4}{9}+\frac{3}{2}\text{III}_{\mathbf{a}}$};
        \node[black] at (-1, -1.9) {\footnotesize 3C};
        \node[black] at (-1.7, -0.75) {\footnotesize 2C};
        \node[black] at (1.9, 2.25) {\footnotesize 1C};
    \end{tikzpicture}
    \caption{\gls{AIM} of the Reynolds stress tensor \newText{comparing second and third invariant of respective anisotropy tensor}. \newText{The corners of the triangle (1C, 2C,3C) represent the componentiality of turbulence (see \cref{tab:turbulentLimitingStates})}\cancelText{ highlighting its limiting states}.}
    \label{fig:invariantAnisotropyPlot}
\end{figure}

\noindent Generally, the anisotropy tensor describes and measures the deviation of the Reynolds stress tensor from the isotropic state, where its geometric \newText{ellipsoid} representation forms a perfect sphere ($\rho_1=\rho_2 = \rho_3$). 
The invariants of the anisotropy tensor
\begin{equation}
\label{eq:invariants}
    \begin{split}
		\text{I}_{\mathbf{a}} &= \text{tr}\left(\mathbf{a}\right) = 0 \\
        \text{II}_{\mathbf{a}} &= -\frac{1}{2}\text{tr}\left(\mathbf{a}^2\right)=\lambda_1\lambda_2+\lambda_1\lambda_3+\lambda_2\lambda_3 \\
        \text{III}_{\mathbf{a}} &= \det\left(\mathbf{a}\right)=\lambda_1\lambda_2\lambda_3
    \end{split}
\end{equation} 
can be used to visualize the tensor in a coordinate-system-invariant way, called the \gls{AIM}~\cite{Lumley}, in \cref{fig:invariantAnisotropyPlot}.

\noindent Because of the physical realizability constraints of the Reynolds stress  tensor~\cite{Schumann1977}
\begin{equation}
\tau_{\alpha \alpha} \geq 0 \ \text{,} \quad 
\tau_{\alpha \alpha} \cdot \tau_{\beta \beta} \geq \tau_{\alpha\beta}^2 \ \text{,} \quad
\text{det}\left(\mathbf{\tau}\right) \geq 0 \ \text{,} \ 
\quad \alpha, \beta ={1,2,3}
\label{eq:schumannRealizable}
\end{equation}
and the definition of the anisotropy tensor (see \cref{eq:decompositionTau}), the entries of the anisotropy tensor are bounded in the following ranges:
\begin{equation}
\label{eq:limitsAnisotropy}
\begin{split}
a_{ij} \in \left\{
\begin{array}{ll} \left[-\frac{2}{3}, \frac{4}{3}\right] \ &\text{for} \ i=j \\
         \left[-1, 1\right] \ &\text{for} \ i \neq j
         \end{array}\right. .
         \end{split}
\end{equation}
The eigenspace decomposition of the anisotropy tensor in combination with tensor diagonalization (see \cref{eq:spectralDecompositionAnisotropy}) leads to the fact that any physically realizable Reynolds stress tensor can be mapped to exactly one respective anisotropy tensor in its canonical form $\Lambda_{ij} = \mathrm{diag} (\lambda_1, \lambda_2, \lambda_3)$.
Applying\cancelText{ Section II A} \newText{\cref{eq:limitsAnisotropy}} to $\Lambda_{ij}$, the ordered eigenvalues
\begin{equation}
\label{eq:boundedNessEigenvalues}
    \begin{split}
    \lambda_1 &= \max_\alpha \left(a_{\alpha \alpha}|_\mathrm{PCS}\right)\\
    \lambda_2 &= \max_{\beta \neq \alpha} \left(a_{\beta\beta}|_\mathrm{PCS}\right) \\
     \lambda_3 &= -\lambda_1 - \lambda_2 = \min_{\gamma \neq \alpha,\beta} \left(a_{\gamma\gamma}|_\mathrm{PCS}\right) \ \text{,}
     \end{split}
\end{equation}
are bounded accordingly~\cite{Terentiev2006}:
\begin{equation}
\label{eq:boundedNessEigenvalues2}
    \lambda_1 \geq \frac{3|\lambda_2| - \lambda_2}{2} \ \text{,} \ \lambda_1 \leq \frac{1}{3}-\lambda_2 \ \text{.}
\end{equation}

\cancelText{Besides, t}\newText{T}urbulence componentiality~\cite{Terentiev2006} categorizes three fundamental states (one-, two- and three-component turbulence) based on the \newText{number of} non-zero eigenvalues of the Reynolds stress tensor $\rho_i$ (and respective anisotropy tensor eigenvalues $\lambda_i$), presented in \cref{tab:turbulentLimitingStates}. Besides, axisymmetric turbulence is characterized by two eigenvalues being equal, while an isotropic state features three identical eigenvalues.
\begin{table}
  \caption{\label{tab:turbulentLimitingStates}Turbulence componentiality and limiting states of turbulence with respect to eigenvalues of the Reynolds stress tensor $\rho_i$ and the anisotropy tensor $\lambda_i$.}
  \begin{ruledtabular}
  \begin{tabular}{l c  c c}
    \multirow{2}{*}{States of turbulence}& componentiality & \multicolumn{2}{c}{eigenvalues}\\ 
    & \# $\rho_i \neq 0$  or \#  $\lambda_i \neq -\frac{2}{3}$ & $\rho_i$& $\lambda_i$\\
    \hline
    One-component  (1C)  &1   &$\rho_1 =2k, \rho_2=\rho_3=0$ & $\lambda_1 =\frac{4}{3}, \lambda_2=\lambda_3=-\frac{2}{3}$\\
    Two-component    &2  &$\rho_1+\rho_2=2k, \rho_3=0$ & $\lambda_1+\lambda_2=\frac{2}{3}, \lambda_3=-\frac{2}{3}$  \\
    Two-component axisymmetric (2C)   &2  &$\rho_1=\rho_2=k, \rho_3=0$ & $\lambda_1=\lambda_2=\frac{1}{3}, \lambda_3=-\frac{2}{3}$  \\
    Three-component    &3 &$\rho_1+\rho_2+\rho_3=2k$ & $\lambda_1+\lambda_2+\lambda_3=0$  \\
    Three-component isotropic (3C)  &3 &$\rho_1=\rho_2=\rho_3=\frac{2}{3}k$ & $\lambda_1=\lambda_2=\lambda_3=0$  \\
    \end{tabular}
  \end{ruledtabular}
\end{table}
\noindent The corners of the \gls{AIM} in \cref{fig:invariantAnisotropyPlot} can be\cancelText{ identified} \newText{classified} as the (three-component) isotropic limit (3C), the two-component axisymmetric limit (2C) and the one-component limit (1C) (see also \cref{tab:turbulentLimitingStates}). Moreover, due to the boundedness of the anisotropy tensor entries (and its eigenvalues, respectively), all physically plausible\cancelText{ turbulence states} \newText{states of turbulence} must lie within the area spanned by the corners of the triangle.
Furthermore, due to the boundedness of the anisotropy tensor\newText{'} eigenvalues, a barycentric triangle\cancelText{ based on the spectral theorem} can be constructed \newText{based on the spectral theorem}~\cite{Banerjee2007}. Consequently, every physically realizable state of the Reynolds stress tensor can be mapped onto barycentric coordinates
\begin{equation}
\label{eq:barycentricMapping}
    \begin{split}
	   \mathbf{x} =& \frac{1}{2}\mathbf{x}_{\mathrm{1C}}\left(\lambda_1-\lambda_2\right)+ \mathbf{x}_{\mathrm{2C}}\left(\lambda_2-\lambda_3\right)+ \mathbf{x}_{\mathrm{3C}} \left(\frac{3}{2}  \lambda_3+1\right)  \\
        \mathbf{x} =& \mathbf{Q} \boldsymbol{\lambda} 
        \quad \text{with} \ \lambda_1\geq\lambda_2 \geq\lambda_3 \ \text{,}
     \end{split}
\end{equation}
where $\mathbf{Q}$ depends on the choice of corners of the barycentric triangle.
\cref{fig:baryCentricTriangle} shows these three limiting states of the Reynolds stress tensor, defined by the corners of the triangle ($\mathbf{x}_{\mathrm{1C}}, \mathbf{x}_{\mathrm{2C}}, \mathbf{x}_{\mathrm{3C}}$) representing the one-component, two-component axisymmetric and three-component (isotropic) turbulent state.
A great benefit of the \gls{ABM} is the possibility to obtain a linear interpolation between two points \newText{with respect to their eigenvalues}.
\newText{The eigenspace perturbation exploits this property as well.}\cancelText{ This is exploited by the eigenspace perturbation as well.} \newText{Hence, we}\cancelText{ We} will come back to it later.

\subsection{\label{sec:perturbation}Perturbation of Eigenspace Representation}

\cancelText{When} \newText{As} the Reynolds\cancelText{ S} \newText{s}tresses are expressed as functions of the mean flow quantities for turbulence modeling, we need to consider the nature of their relationship. A common example are the state-of-the-art \gls{LEVM}, which assume this relationship to be linear and introduce a turbulent (eddy) viscosity $\nu_t$ to approximate the Reynolds stress tensor in analogy to the viscous stresses
\begin{equation}
\label{eq:boussinesq}
    \tau_{ij} = -2 \nu_t \left(S_{ij} - \frac{1}{3}\frac{\partial u_k}{\partial x_k}\delta_{ij}\right) + \frac{2}{3} k \delta_{ij} \ \text{,}
\end{equation}
where the strain-rate tensor is denoted as $S_{ij}$. In the past decades researchers have pointed out limitations of these \gls{LEVM} for flow situations, which are not covered by the calibration cases~\cite{Speziale, Mompean, CRAFT1996108, Lien}.
The estimated relationship between Reynolds stresses and mean rate of strain results in the inability to account correctly for its anisotropy and consequently lead to a significant degree of epistemic uncertainty. 
In order to account for\cancelText{ the} such epistemic uncertainties due to the model-form, the perturbation approach suggests to modify the eigenspace (eigenvalues and eigenvectors) of the Reynolds stress tensor within physically permissible limits~\cite{Emory, Iaccarino}.
The \gls{EPF} of the Reynolds stress tensor implemented in  \textit{TRACE} creates a perturbed state of the Reynolds stress tensor defined as
\begin{equation}
	\label{eq:spectralDecompositionR*}
	\begin{split}
	    \tau_{ij}^* &= k \left(a_{ij}^* + \frac{2}{3}\delta_{ij}\right) \\
		            &= k \left(v_{in}^* \Lambda_{nl}^* v_{jl}^* + \frac{2}{3}\delta_{ij}\right) \ \text{,}
	\end{split}
\end{equation}
where $a_{ij}^*$ is the perturbed anisotropy tensor, $\Lambda_{nl}^*$ is\cancelText{ the} the perturbed eigenvalue matrix and $v^*_{in}$ is the perturbed eigenvector matrix. The turbulent kinetic $k$ energy is left unchanged.
In the following sections, we will describe the mathematical and physical foundation of forming a perturbed eigenspace.

\subsubsection{\label{sec:eigenvaluePerturbation} Eigenvalue perturbation}
\begin{figure}
\centering
\begin{tikzpicture}
\node[] at (0,3) {
                \includegraphics[width=0.07\textwidth]{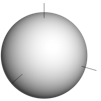}};
\node[] at (-2,-2.6) {
                \includegraphics[width=0.07\textwidth]{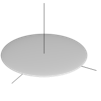}};
\node[] at (2,-2.6) {
                \includegraphics[width=0.07\textwidth]{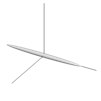}};
\draw [-](-2,-1.5) -- (2,-1.5) node[] {};
\draw [-](-2,-1.5) -- (0,1.964) node[] {};
\draw [-]( 2,-1.5) -- (0,1.964) node[] {};
\draw [gray, dashed]( -0.3, 0.5) -- (0,1.964) node[] {};
\draw [gray, dashed]( -1.34, -0.72) -- (-2,-1.5) node[] {};
\draw [gray, dashed]( -0.3, 0.5) -- (0.2,0.08) node[] {};
\draw [gray, dashed]( 0.82, -0.48) -- (2,-1.5) node[] {};
\draw [red]( -1.34, -0.72) -- (-0.3, 0.5) node[] {};
\filldraw[black] (-0.3, 0.5) circle (2pt) node[anchor=west]{$\mathbf{x}$};
\filldraw[black] (1.7, -1.5) circle (0.01pt) node[anchor=north west]{$\mathbf{x}_{\mathrm{1C}}$};
\filldraw[black] (-1.7, -1.5) circle (0.01pt) node[anchor=north east]{$\mathbf{x}_{\mathrm{2C}}$};
\filldraw[black] (0, 1.964) circle (0.01pt) node[anchor=south]{$\mathbf{x}_{\mathrm{3C}}$};
\filldraw[black] ( -1.34, -0.72) circle (2pt) node[anchor=west]{$\mathbf{x}^*$};
\node[red] at (0, -0.3) {$\Delta_B = \frac{\mathbf{x}^* - \mathbf{x}}{\mathbf{x}_{\mathrm{2C}} - \mathbf{x}}$};
\node[black] at (0, -1.7) {\footnotesize Two-component};
\node[black, rotate=60] at (-1.19, 0.3) {\footnotesize Axisymmetric};
\node[black, rotate=-60] at (1.19, 0.3) {\footnotesize Axisymmetric};
\end{tikzpicture}
\caption{\gls{ABM} representing the eigenvalues of the anisotropy tensor and its effect on the shape of the Reynolds stress tensor ellipsoid. The eigenvalue perturbation towards the two-component limiting state of turbulence is shown schematically.}
\label{fig:baryCentricTriangle}
\end{figure}
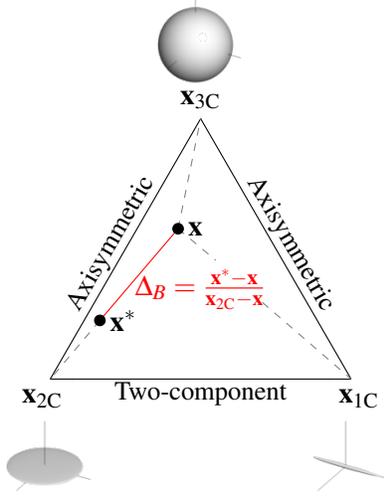

The eigenvalue perturbation utilizes the boundedness of the \newText{eigenvalues of the} anisotropy tensor\cancelText{ eigenvalues} and their representation in terms of barycentric coordinates, as described in \cref{sec:eigdecomposition}. 
As the representation of the anisotropy tensor within the \gls{ABM} enables linear interpolation between a starting point $\mathbf{x}$ and a target point $\mathbf{x}_{(t)}$, the perturbation methods creates a modified location $\mathbf{x}^*$, according to 
\begin{equation}
	\label{eq:perturbationMagnitude}
		\mathbf{x}^* = \mathbf{x} + \Delta_B \left(\mathbf{x}_{(t)} -\mathbf{x}\right) \ \text{,}
\end{equation}
with the relative distance $\Delta_B \in [0, 1]$ controlling the magnitude of eigenvalue perturbation as illustrated in \cref{fig:baryCentricTriangle}. The starting point $\mathbf{x}$ is usually determined in the \gls{RANS} simulation iteration via the relationship for the Reynolds stresses determined by the turbulence model, e.g. the\cancelText{ Boussinseq} \newText{Boussinesq} assumption for \gls{LEVM} (see \cref{eq:boussinesq}).
Due to their distinctive significance, the limiting states of turbulence at the corners\cancelText{ typically} act \newText{typically} as the target point $\mathbf{x}_{(t)} \in \{\mathbf{x}_{\mathrm{1C}}, \mathbf{x}_{\mathrm{2C}}, \mathbf{x}_{\mathrm{3C}}\}$.
Subsequently, the perturbed eigenvalues $\lambda_{i}^*$ can be remapped by the inverse of $\mathbf{Q}$
\begin{equation}
	\label{eq:perturbedEigenvalues}
		\boldsymbol{\lambda}^* = \mathbf{Q}^{-1} \mathbf{x}^* \ \text{.}
\end{equation}

\subsubsection{\label{sec:eigenvectorPerturbation} Eigenvector perturbation}
In contrast to the eigenvalues, there are no physical bounds for the orientation of the eigenvectors of the Reynolds stress tensor and there is no upper limit for the turbulent kinetic energy. Thus, the fundamental idea of perturbing the eigenvectors is to create bounding states for the production $P_k$ of turbulent kinetic energy $k$ in\cancelText{ a} transport equation based \gls{LEVM}\cancelText{ context}. Hereby, the budget of turbulent kinetic energy is indirectly manipulated. 
The turbulent production term is defined as the Frobenius inner product of the Reynolds stress and the strain-rate tensor\cancelText{ s}. Since both are positive semi-definite, the bounds of the Frobenius inner product can be written in terms of their eigenvalues $\rho_i$ and $\sigma_i$ arranged in decreasing order~\cite{Lasserre}:
\begin{equation}
\label{eq:turbulentProduction}
    \begin{split}
        P_k &=-\tau_{ij}\frac{\partial u_i}{\partial x_j} 
            = -\tau_{ij} \cdot S_{ij} 
            =-\langle \boldsymbol{\tau}, \mathbf{S}\rangle_F 
            = -\text{tr}\left(\boldsymbol{\tau} \mathbf{S}\right) \\
            &\in \left[\rho_1\sigma_3+\rho_2\sigma_2+\rho_3\sigma_1, \, \rho_1\sigma_1+\rho_2\sigma_2+\rho_3\sigma_3\right]  \ \text{.}
        \end{split}
\end{equation}
Since the Reynolds stress\cancelText{ tensor} and the strain rate tensor share the same eigenvectors in \gls{LEVM} \newText{(see \cref{eq:boussinesq})}, the lower bound of the turbulent production term can be obtained by commuting the first and third eigenvector of the Reynolds stress tensor, whereas maximum turbulent production is obtained by not changing the eigenvectors of the Reynolds stress tensor:
\begin{equation}
    \begin{split}
        \label{eq:minMaxProduction}
        \mathbf{v}_\textrm{max} &= 
        \begin{pmatrix}
         \mathbf{v}_{1_{\mathbf{S}}} & \mathbf{v}_{2_{\mathbf{S}}} & \mathbf{v}_{3_{\mathbf{S}}} 
        \end{pmatrix} \rightarrow P_{k_\textrm{max}} \\
        \mathbf{v}_\textrm{min} &= 
        \begin{pmatrix}
        \mathbf{v}_{3_{\mathbf{S}}} & \mathbf{v}_{2_{\mathbf{S}}} & \mathbf{v}_{1_{\mathbf{S}}} 
        \end{pmatrix} \rightarrow P_{k_\textrm{min}} \ \text{.}
    \end{split}
\end{equation}
Note: Permuting of the eigenvectors of the Reynolds stress is equivalent to changing the order of the respective eigenvalues. Both change the alignment of the Reynolds stress ellipsoid with the principle axes of the strain-rate tensor.

\subsubsection{\label{sec:practivalusage} Implications for \gls{CFD} practitioners}
\cancelText{In Summary, t} \newText{T}he eigenspace perturbation can be divided into eigenvalue and eigenvector modifications of the Reynolds stress tensor.
For practical application purposes each eigenvalue perturbation towards one of the limiting states of turbulence\cancelText{ is} \newText{can be} combined with\cancelText{ two approaches aiming to minimize and maximize} \newText{minimization or maximization of} the turbulent production term (eigenvector perturbation)\cancelText{ , respectively}. 
In summary, the model-form uncertainty of\cancelText{ a} \gls{LEVM} can be estimated by\cancelText{ at least 5 additional \gls{CFD} simulations if $\Delta_B = 1$ is chosen and 6 perturbed simulations if $\Delta_B < 1$} \newText{6 additional \gls{CFD} simulations if $\Delta_B < 1$ and only 5 perturbed simulations if $\Delta_B = 1$ is chosen. This is because the Reynolds stress ellipsoid is a perfect sphere when targeting for the $\mathrm{3C}$ turbulence state with $\Delta_B =1$ (see \cref{fig:baryCentricTriangle}), making an eigenvector perturbation obsolete}.\cancelText{ For the 3C turbulent target state with $\Delta_B =1$, the Reynolds stress ellipsoid is a perfect sphere (see \cref{fig:baryCentricTriangle}), so no eigenvector perturbation is necessary in this case.} As the amount of \newText{considered} turbulence model uncertainty\cancelText{ considered} scales with the relative perturbation strength $\Delta_B$, aiming for the corners of the barycentric triangle (applying $\Delta_B = 1$) is common practice in order to obtain a worst case estimate corresponding to the most conservative uncertainty bounds on \gls{QoI} ~\cite{Emory, Iaccarino, MishraSU2, MathaCF}.
The analysis of\cancelText{ the} additional\cancelText{ perturbed} \gls{CFD} simulations, \newText{propagating the effect of perturbed Reynolds stress tensor,} enables a \gls{CFD} practitioner to quantify the derived effect of the turbulence model perturbation on certain \gls{QoI}, e.g. the pressure field.

\subsection{\label{sec:self-consistentFomrulation}Self-consistent formulation of perturbation}
The emergence of\cancelText{ the} \newText{some} shortcomings of the eigenspace perturbation of \newText{the} Reynolds stress tensor is highlighted in this section.\cancelText{ Based on the} \newText{This forms the foundation of rethinking of the computational formulation the \gls{EPF}}. The present paper suggests an appropriate way of formulating the \gls{EPF}, ensuring control over numerical stability while preserving the conceptual model of perturbing the eigenspace of the Reynolds stress tensor.

\subsubsection{\label{sec:needForModeration}Need for moderating the perturbation strength}
The need for moderating the effect of Reynolds stress tensor perturbation emerges, when the Reynolds stress tensor perturbation seeks to decrease the turbulent kinetic energy budget ($P_{k_\textrm{min}}$ and/or $\mathrm{3C}$). These perturbations featuring overly reduced turbulent viscosity can lead to numerical convergence issues for example when simulating separated flows. 
To ensure convergence while still perturbing as much as required, there is a need to moderate the effect of Reynolds stress tensor perturbation. \newText{Recent publications introduce}\cancelText{ The introduction of} a moderation factor $f$ \newText{to enable}\cancelText{ enables} the \gls{CFD}-solver to achieve fully converged, steady-state \gls{RANS} results~\cite{MishraSU2, MathaCF}. 
Consequently, the \newText{propagated} perturbed Reynolds stress tensor (entering the update of the viscous fluxes and the turbulent production term) can be expressed as
\begin{equation}
    \begin{split}
        \label{eq:moderationFactorReconstruction}
        \tau_{{ij}_f}^* &= \tau_{ij} + f \left[\tau_{ij}^* - \tau_{ij}\right]\text{,}
    \end{split}
\end{equation}
where $f \in [0, 1]$ is the introduced moderation factor, adjusting the total amount of perturbed anisotropy tensor to be considered.
Note: The effect of applying the moderation factor is identical to a reduction of $\Delta_B$ in \cref{eq:perturbationMagnitude} in case of pure eigenvalue perturbation~\cite{MathaCF}.

\subsubsection{\label{sec:unwantedEffect}Inconsistency when combining eigenspace perturbation and moderation factor}

Unfortunately, the unperturbed Reynolds stress tensor $\tau_{ij}$ and the perturbed one $\tau_{ij}^*$ do not necessarily share the same eigenvectors. When eigenvector perturbation is applied, the resulting moderated Reynolds stress tensor $\tau_{{ij}_f}^*$ shows unintended behaviour with respect to its projection onto barycentric coordinates.
\begin{figure}
    \begin{tikzpicture}
        \node[] at (0,0) {
                \includegraphics[width=0.35\textwidth, trim=0.7cm 0.5cm 0.7cm 6cm, clip=True]{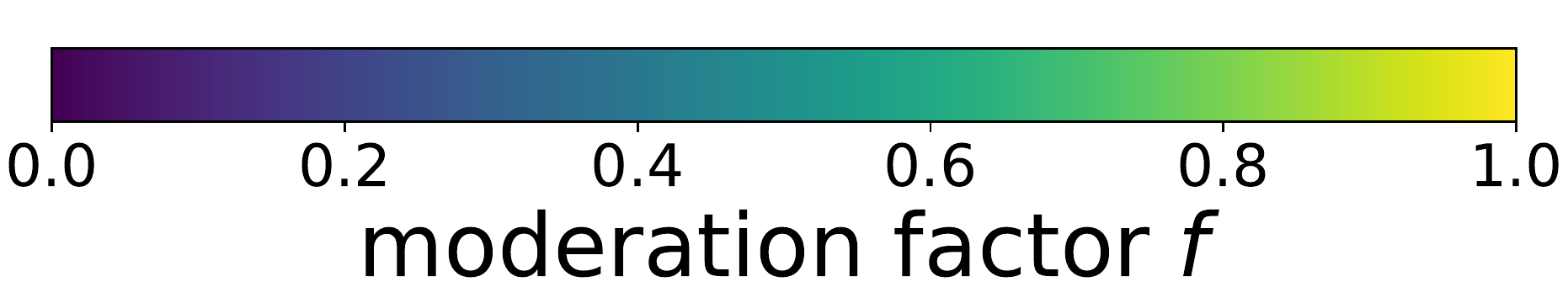}};
    \end{tikzpicture}\\
    \begin{subfloat}[\label{fig:baryCoordChannelFlowWOeigenvectors}]
         {\includegraphics[width=0.3\textwidth, trim=0.9cm 1cm 0.7cm 1.1cm, clip=True]{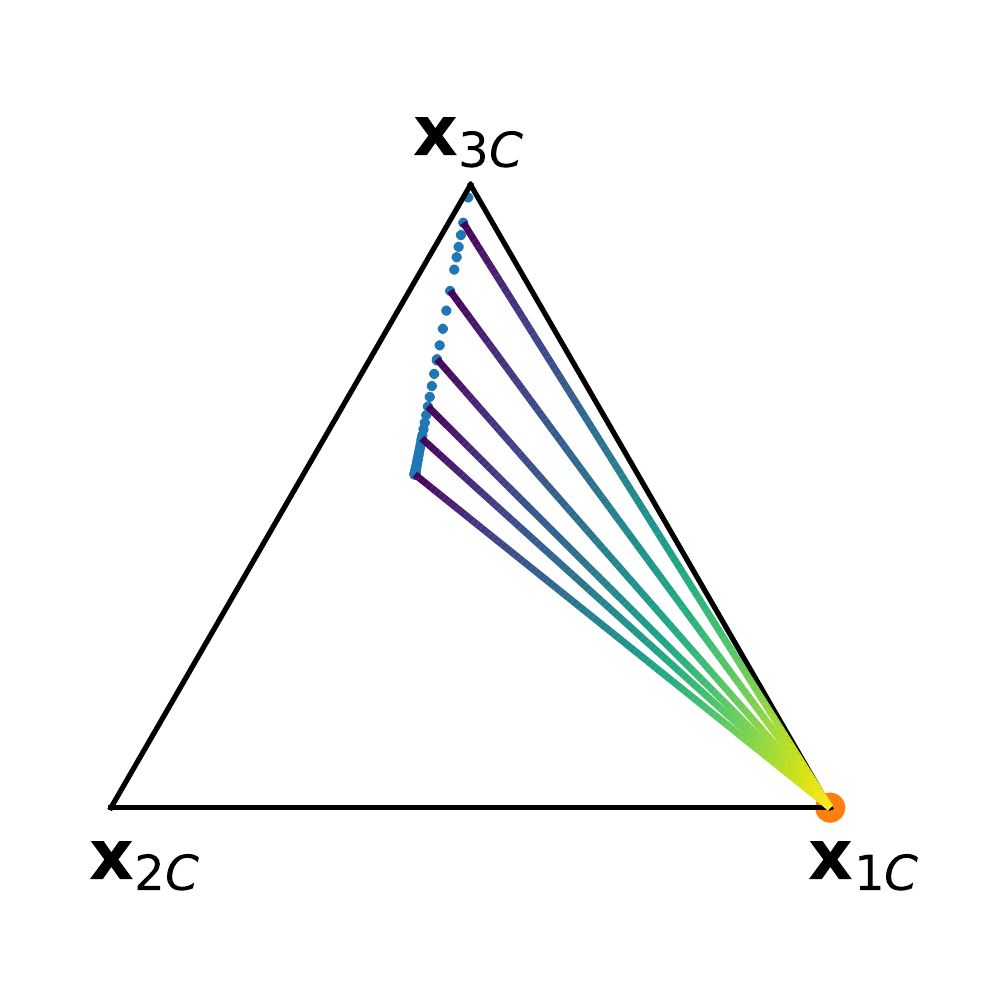}}
     \end{subfloat}
     \begin{subfloat}[\label{fig:baryCoordChannelFlowWEigenvectors}]
         {\includegraphics[width=0.3\textwidth, trim=0.7cm 1cm 0.9cm 1.1cm, clip=True]{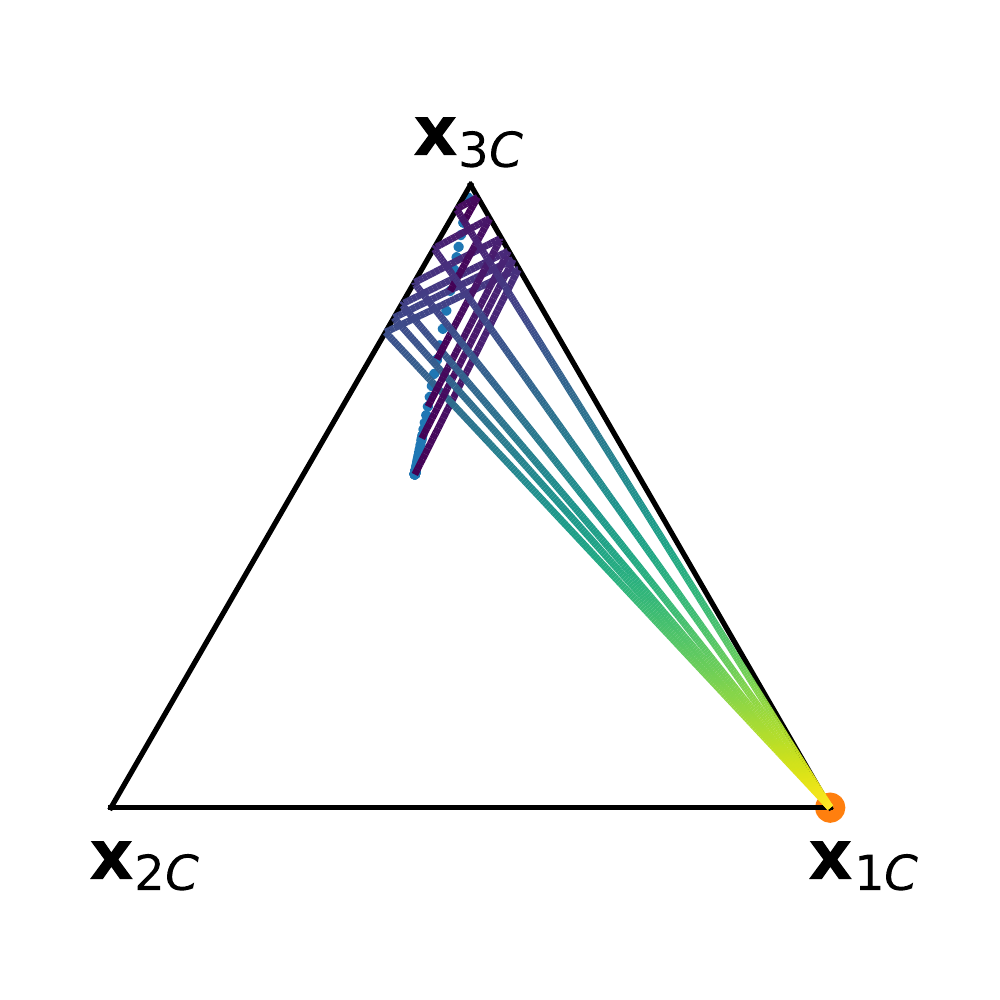}}
      \end{subfloat}
\caption{Comparison of perturbation trajectory for \gls{RANS} channel flow data at $Re_\tau = 1000$ (blue dots) in barycentric coordinates. The trajectories for selected \gls{RANS} data points ($\tau_{ij}$) are created by increasing $f=0...1$ with and without eigenvector perturbation targeting the one-component limiting state of turbulence $\tau_{ij}^* = \tau_{{ij}_{\mathrm{1C}}}$ (orange dot). (a) Without eigenvector modification (aiming for $P_{k_\textrm{max}}$). (b) Perturbation including eigenvector modification (aiming for $P_{k_\textrm{min}}$).}
\label{fig:channelFlowExample}
\end{figure}
\cref{fig:channelFlowExample} presents the perturbation trajectory\cancelText{ by} \newText{when}  increasing $f$ from $0$ to $1$ for selected \gls{RANS} data points inside the \gls{AIM} towards the one-component limiting state of turbulence. 
 The moderated Reynolds stress tensor is calculated based on \cref{eq:moderationFactorReconstruction} with $\tau_{ij}^* = \tau_{{ij}_{\mathrm{1C}}}$, while $\tau_{{ij}_{\mathrm{1C}}}$ is a function of $\Lambda^*_{{ij}_{\mathrm{1C}}}$, $\mathbf{v}^*_i$ and $k_{\mathrm{RANS}}$. Each location along the perturbation trajectory results from determining the respective moderated anisotropy tensor and its barycentric coordinates related to its eigenvalues.
The perturbation trajectory when $\tau_{ij}^*$ and $\tau_{ij}$ share identical eigenvectors shows the expected linear interpolation between the respective coordinates. However, when applying eigenvector perturbation (first and last column of $v^*_{jl}$ are commuted) the resulting intermediate paths do not represent the most direct connection between starting and target point.
Instead, the perturbation trajectories in \cref{fig:baryCoordChannelFlowWEigenvectors} point towards axisymmetric expansion (line between $\mathbf{x}_{\mathrm{3C}}$ and $\mathbf{x}_{\mathrm{1C}}$) first,\cancelText{ then} \newText{head} towards axisymmetric contraction (line between $\mathbf{x}_{\mathrm{3C}}$ and $\mathbf{x}_{\mathrm{2C}}$) \newText{subsequently} and\cancelText{ finally} target the one-component limit of turbulence \newText{finally}.\\
The mathematical explanation for this observation\cancelText{ on channel flow data}, when combining eigenvalue and eigenvector perturbation\cancelText{ and} \newText{while} moderating their effect\newText{s} by a factor according to \cref{eq:moderationFactorReconstruction} is given\cancelText{ subsequently} \newText{thereupon}.
Thus, the\cancelText{ preconditions} \newText{prerequisites} for the accomplishment of linear interpolation properties in terms of barycentric coordinates, when adding two tensors $\mathbf{X}$ and $\mathbf{Y}$, are addressed.
Assuming $\mathbf{X}$ and $\mathbf{Y}$ are positive semi-definite (as the Reynolds stress tensor), then these tensors are realizable~\cite{Schumann1977} and their projection onto barycentric coordinates has to lie\cancelText{ inside} \newText{within} the barycentric triangle~\cite{Banerjee2007}, following the reasons mentioned above (see \cref{sec:eigdecomposition}).
If $\mathbf{X}$ and $\mathbf{Y}$ share identical eigenvectors (commuting matrices), their sum $\mathbf{X}+\mathbf{Y}$ will feature the same eigenvectors and its eigenvalues are the sum of the individual eigenvalues of $\mathbf{X}$ and $\mathbf{Y}$ consequently (see \cref{app:eigenvectors}).
Moreover, if $\mathbf{X}$ and $\mathbf{Y}$ are positive semi-definite, their sum $\mathbf{X}+\mathbf{Y}$ will be positive semi-definite as well (see \cref{app:positiveSemiDefinite}). This implies, that the sum of two realizable Reynolds stress tensors will fulfill realizability constraints and will be located inside the \gls{ABM} accordingly.\\
The line of argument mentioned above is also true for the summation of two scaled tensors 
\begin{equation}
        \label{eq:summationWithFactor}
        \mathbf{Z}  = \mathbf{X} + f \left[\mathbf{Y} - \mathbf{X}\right]
                     =\left(1-f\right)\mathbf{X} + f \mathbf{Y} \ \text{,}
\end{equation}
as multiplying a tensor\cancelText{ with} \newText{by} a scalar does not affect the eigenvectors and modifies the eigenvalues linearly.
The individual scaling of the tensors is chosen, such that the first invariant of $\mathbf{Z}$ ($\text{tr}\left(\mathbf{Z}\right)$) remains identical to $\mathbf{X}$ and $\mathbf{Y}$ ($\text{tr}\left(\mathbf{X}\right)= \text{tr}\left(\mathbf{Y}\right)=\text{tr}\left(\mathbf{Z}\right)$). Keeping in mind, that $\mathbf{X}$ and $\mathbf{Y}$ represent Reynolds stress tensors, this means, that the turbulent kinetic energy remains constant. This is achieved by choosing $f \in [0, 1]$.
Due to the affine transformation, the barycentric coordinates of the anisotropic part of $\mathbf{Z}$ are determined by $\mathbf{x}_\mathbf{Z} = \left(1-f\right)\mathbf{x}_\mathbf{X} + f \mathbf{x}_\mathbf{Y}$, when $\mathbf{x}_\mathbf{X}$ and $\mathbf{x}_\mathbf{Y}$ are the initial states of the tensors $\mathbf{X}$ and $\mathbf{Y}$ in barycentric coordinates (see \cref{app:interpolationProperties}).
\begin{figure}
    \begin{minipage}[b][19cm][c]{0.39\textwidth}
        \begin{tikzpicture}
            \node[] at (0,0) {
                \includegraphics[width=0.85\textwidth, trim=0.7cm 0.5cm 0.7cm 6cm, clip=True]{figures/colorbar2.pdf}};
        \end{tikzpicture}
        \vfil
        \hspace*{0.2cm}
        \begin{subfloat}[\label{fig:baryAtoC}Representation in \gls{ABM}~\cite{Banerjee2007}]{%
        \includegraphics[width=\textwidth, trim=0.cm 1.1cm 0cm 1cm, clip=True]{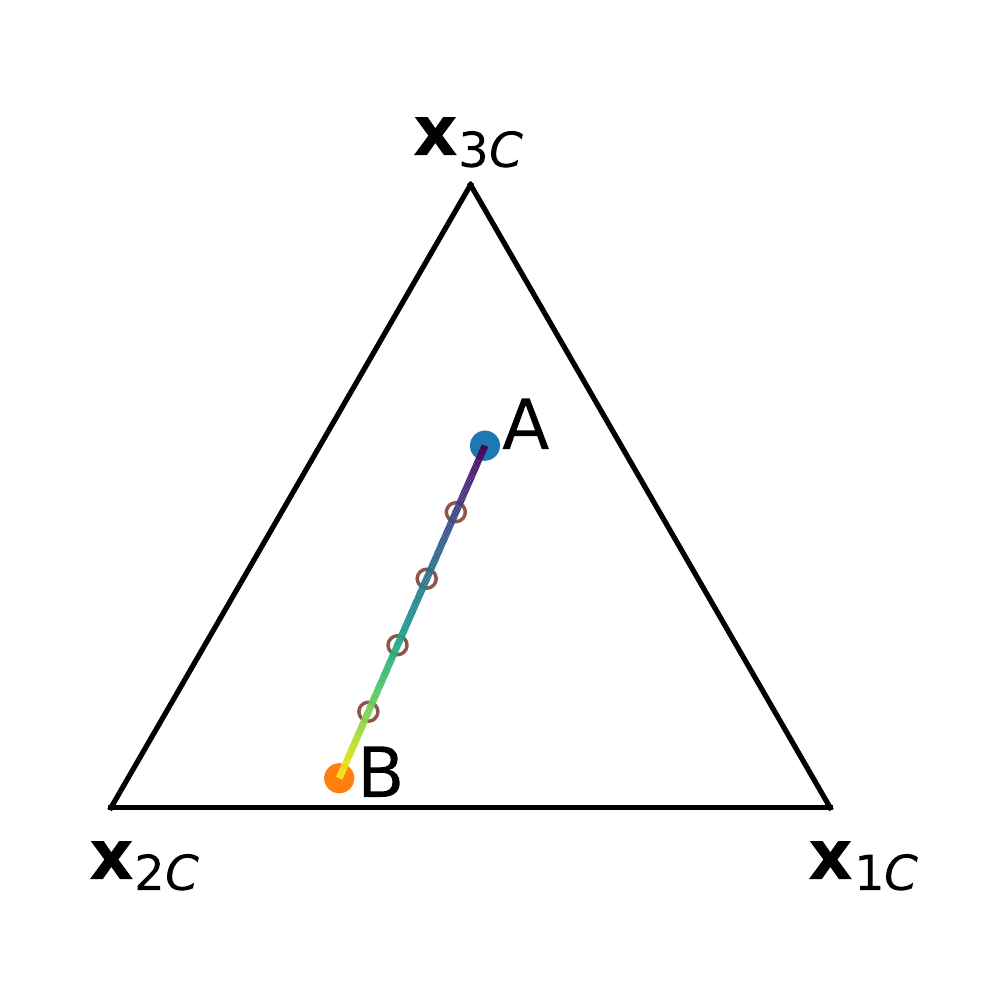}} 
        \vspace{-0.6\baselineskip}
         \end{subfloat}
         \vfil
         \begin{subfloat}[\label{fig:invariantsAtoC}Representation in \gls{AIM}~\cite{Lumley}]
             {\includegraphics[width=\textwidth, trim=0.cm 0.9cm 0cm 0cm, clip=True]{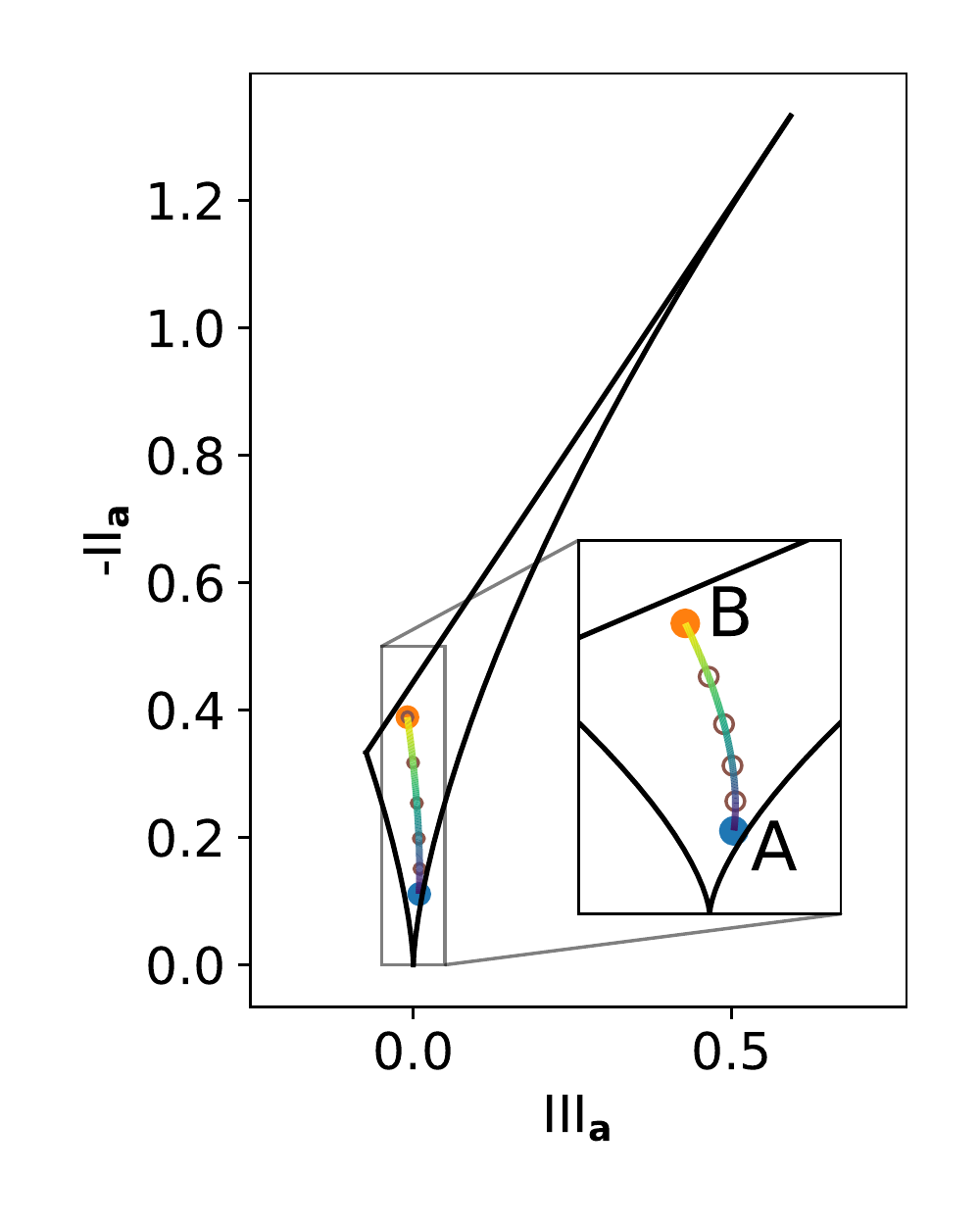}}
             \vspace{-0.6\baselineskip}
         \end{subfloat}
         \vfil
         \begin{subfloat}[\label{fig:xiEtaAtoC}Representation in alternative Anisotropy Invariant Map~\cite{Choi}]
             {\includegraphics[width=\textwidth, trim=0.cm 0.9cm 0cm 0cm, clip=True]{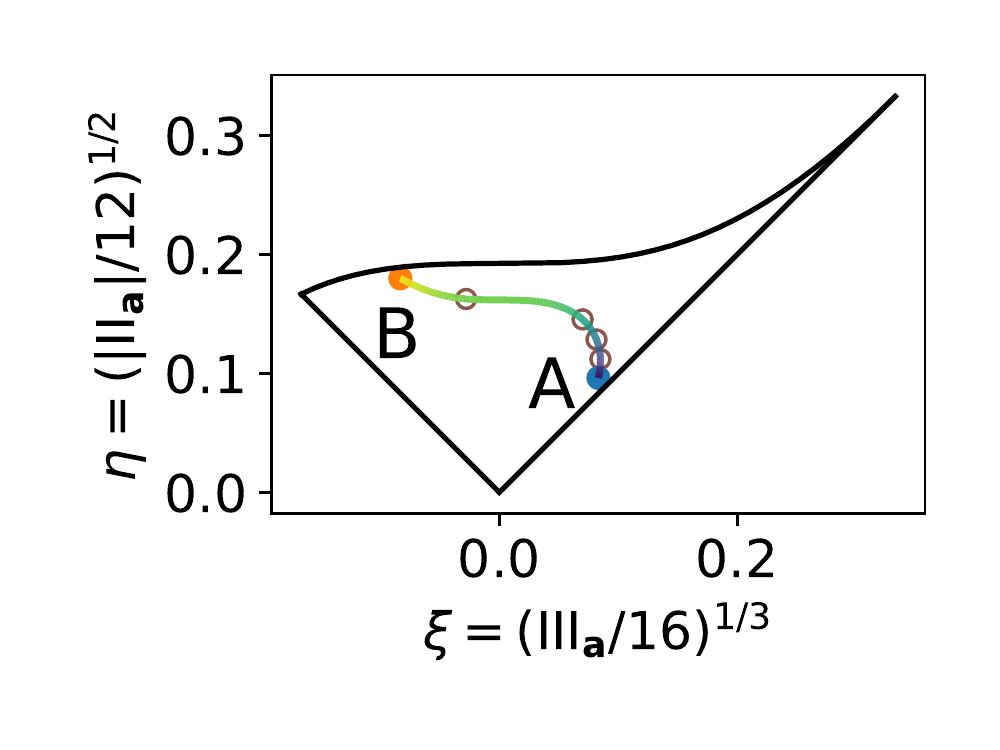}}
             \vspace{-0.6\baselineskip}
         \end{subfloat}
    \end{minipage}
    \begin{subfloat}[\label{fig:tensorEllipsoidAtoC}Reynolds stress tensor ellipsoid]
        {\begin{tikzpicture}
            \node[] at (0,16) {
                \includegraphics[width=0.24\textwidth, trim=3cm 1.8cm 3cm 2cm, clip=True]{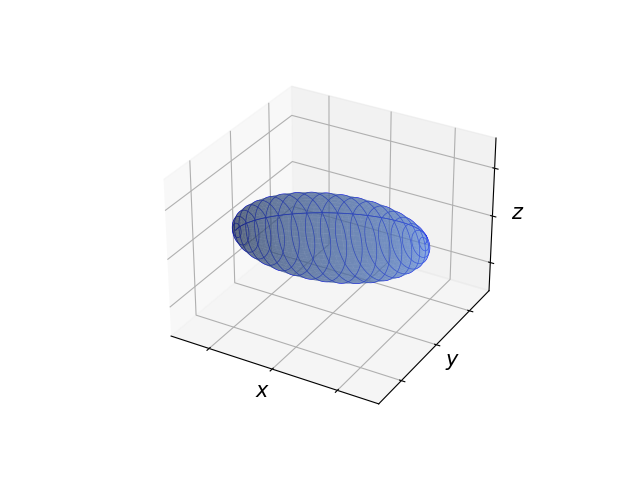}};
            \node[] at (0,12.8) {
                \includegraphics[width=0.24\textwidth, trim=3cm 1.8cm 3cm 2cm, clip=True]{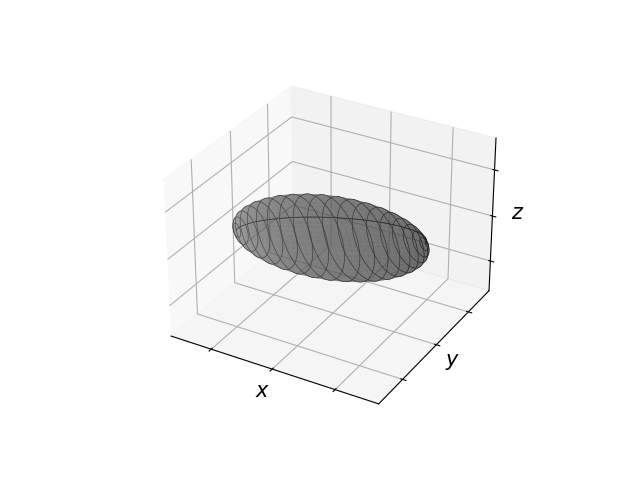}};
            \node[] at (0,9.6) {
                \includegraphics[width=0.24\textwidth, trim=3cm 1.8cm 3cm 2cm, clip=True]{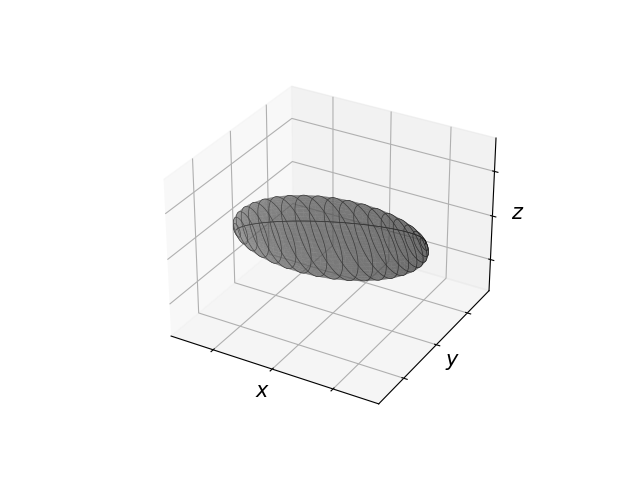}};
            \node[] at (0,6.4) {
                \includegraphics[width=0.24\textwidth, trim=3cm 1.8cm 3cm 2cm, clip=True]{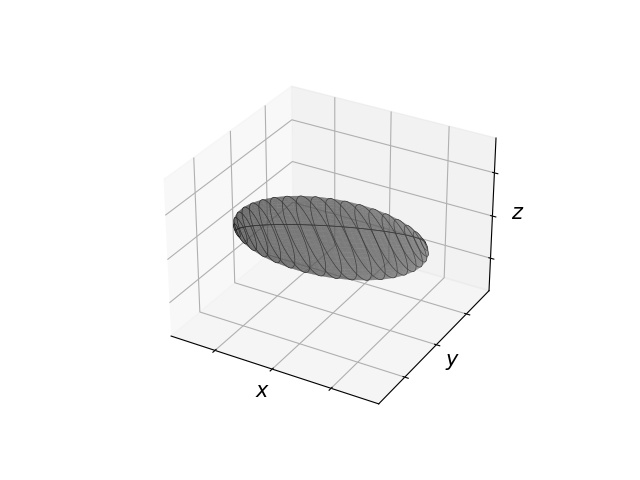}};
            \node[] at (0,3.2) {
                \includegraphics[width=0.24\textwidth, trim=3cm 1.8cm 3cm 2cm, clip=True]{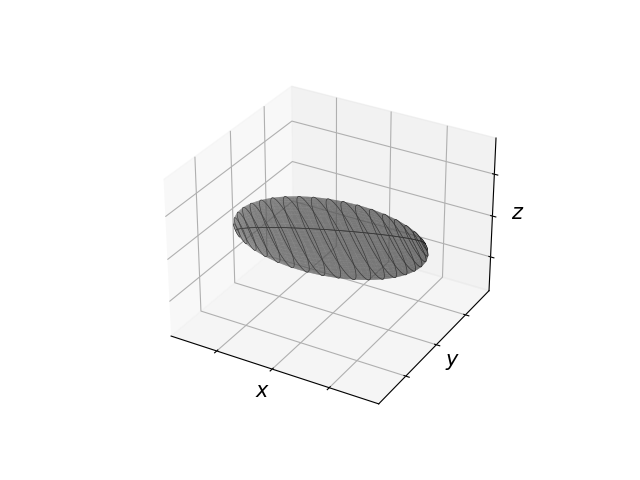}};
            \node[] at (0,0) {
                \includegraphics[width=0.24\textwidth, trim=3cm 1.8cm 3cm 2cm, clip=True]{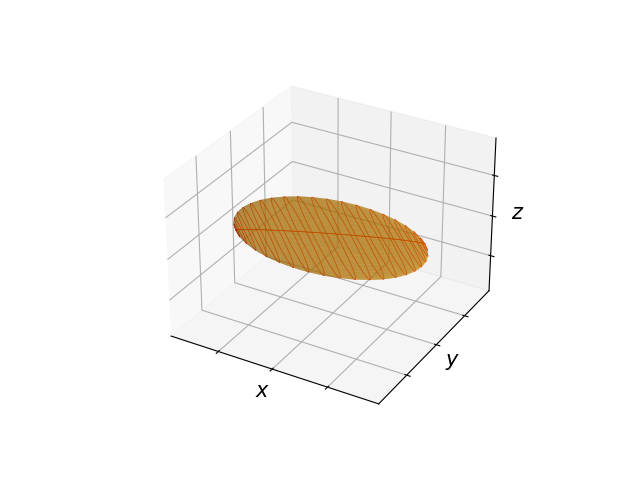}};
            \node[circle, black, draw] at (-1, 17) {$\mathrm{\mathbf{A}}$};
            \node[ black, draw] at (-0.9, 13.6) {\scriptsize $f=0.2$};
            \node[ black, draw] at (-0.9, 10.7) {\scriptsize $f=0.4$};
            \node[ black, draw] at (-0.9, 7.3) {\scriptsize $f=0.6$};
            \node[ black, draw] at (-0.9, 4.0) {\scriptsize $f=0.8$};
            \node[circle, black, draw] at (-1, 0.7) {$\mathrm{\mathbf{B}}$};
        \end{tikzpicture}}  
     \end{subfloat}
     \begin{subfloat}[\label{fig:eigenvectorsAtoC}Eigenvectors of Reynolds stress tensor]
        {\begin{tikzpicture}
            \node[] at (0,16) {
                \includegraphics[width=0.24\textwidth, trim=3cm 1.8cm 3cm 2cm, clip=True]{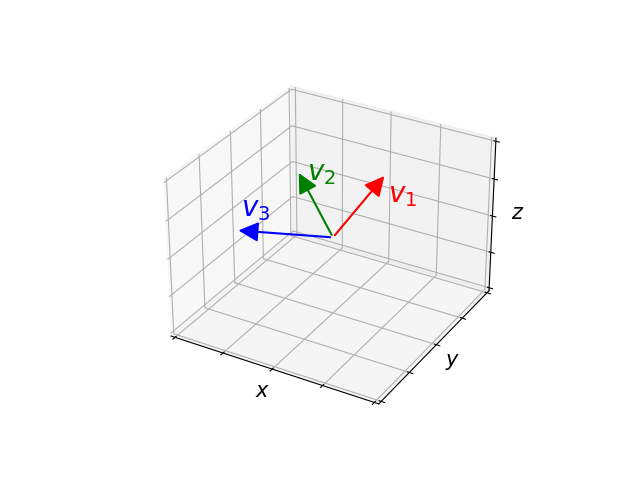}};
            \node[] at (0,12.8) {
                \includegraphics[width=0.24\textwidth, trim=3cm 1.8cm 3cm 2cm, clip=True]{figures/vectorsTransormationAtoBnew4.png}};
            \node[] at (0,9.6) {
                \includegraphics[width=0.24\textwidth, trim=3cm 1.8cm 3cm 2cm, clip=True]{figures/vectorsTransormationAtoBnew4.png}};
            \node[] at (0,6.4) {
                \includegraphics[width=0.24\textwidth, trim=3cm 1.8cm 3cm 2cm, clip=True]{figures/vectorsTransormationAtoBnew4.png}};
            \node[] at (0,3.2) {
                \includegraphics[width=0.24\textwidth, trim=3cm 1.8cm 3cm 2cm, clip=True]{figures/vectorsTransormationAtoBnew4.png}};
            \node[] at (0,0) {
                \includegraphics[width=0.24\textwidth, trim=3cm 1.8cm 3cm 2cm, clip=True]{figures/vectorsTransormationAtoBnew4.png}};
            \node[circle, black, draw] at (-1, 17) {$\mathrm{\mathbf{A}}$};
            \node[ black, draw] at (-0.9, 13.6) {\scriptsize $f=0.2$};
            \node[ black, draw] at (-0.9, 10.7) {\scriptsize $f=0.4$};
            \node[ black, draw] at (-0.9, 7.3) {\scriptsize $f=0.6$};
            \node[ black, draw] at (-0.9, 4.0) {\scriptsize $f=0.8$};
            \node[circle, black, draw] at (-1, 0.7) {$\mathrm{\mathbf{B}}$};
        \end{tikzpicture}}
    \end{subfloat}
\caption{Transition from tensor $\mathrm{\mathbf{A}}$ to $\mathrm{\mathbf{B}}$ (defined in \cref{app:tensorABC}) featuring identical eigenvectors by increasing $f=0...1$ (see  \cref{eq:summationWithFactor}). The intermediate brown-colored states in \protect\subref{fig:baryAtoC}, \protect\subref{fig:invariantsAtoC} and \protect\subref{fig:xiEtaAtoC} correspond to the states with $f\in \text{[}0.2, 0.4, 0.6, 0.8\text{]}$ in \protect\subref{fig:tensorEllipsoidAtoC} and \protect\subref{fig:eigenvectorsAtoC}.}
\label{fig:blendingIdenticalEigenvectors}
\end{figure}
\begin{figure}
    \begin{minipage}[b][19cm][c]{0.39\textwidth}
        \begin{tikzpicture}
            \node[] at (0,0) {
                \includegraphics[width=0.85\textwidth, trim=0.7cm 0.5cm 0.7cm 6cm, clip=True]{figures/colorbar2.pdf}};
        \end{tikzpicture}
        \vfil
        \hspace*{0.2cm}
        \begin{subfloat}[\label{fig:baryAtoB}Representation in \gls{ABM}~\cite{Banerjee2007}]
             {\includegraphics[width=\textwidth, trim=0.cm 1.1cm 0cm 1cm, clip=True]{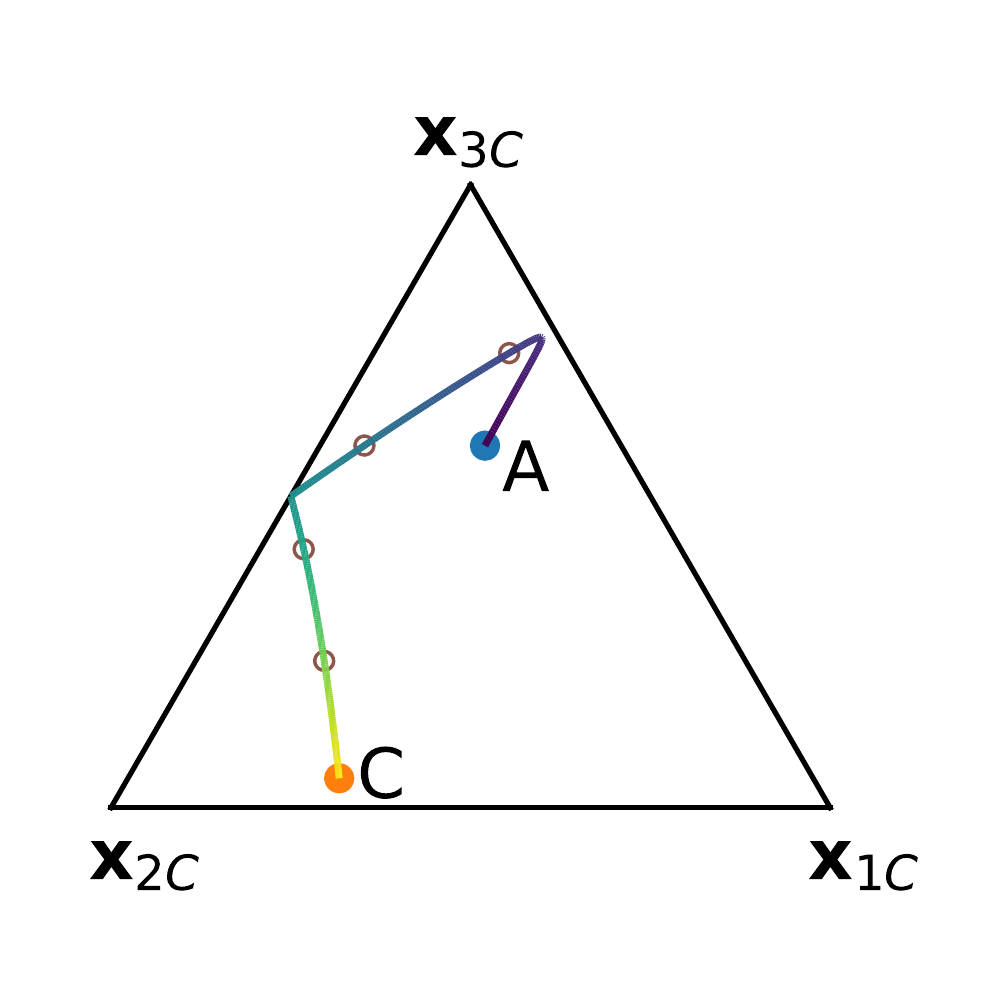}}
             \vspace{-0.6\baselineskip}
         \end{subfloat}
         \vfil
         \begin{subfloat}[\label{fig:invariantsAtoB}Representation in \gls{AIM}~\cite{Lumley}]
             {\includegraphics[width=\textwidth, trim=0.cm 0.9cm 0cm 0cm, clip=True]{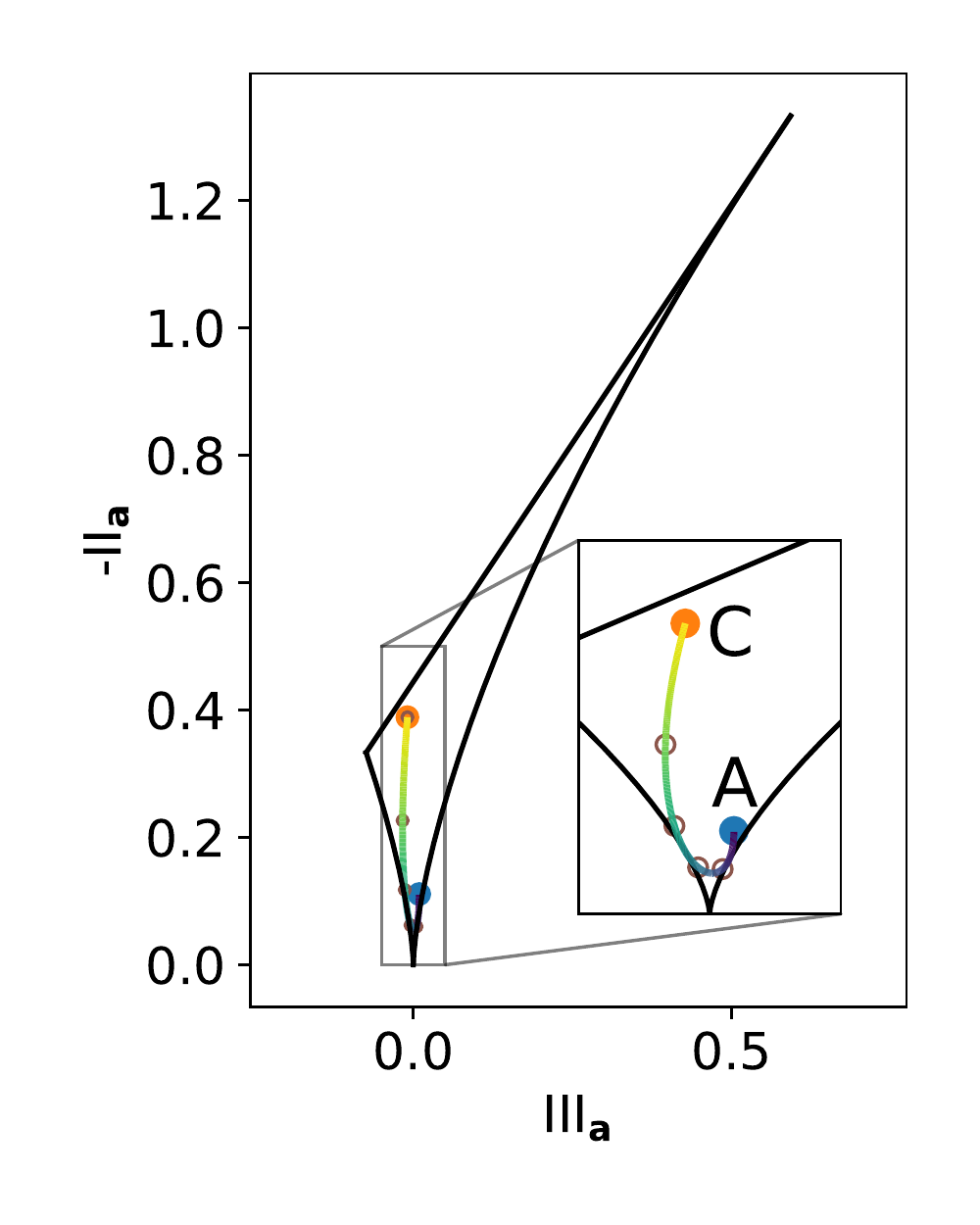}}
             \vspace{-0.6\baselineskip}
         \end{subfloat}
         \vfil
         \begin{subfloat}[\label{fig:xiEtaAtoB}Representation in alternative Anisotropy Invariant Map~\cite{Choi}]
             {\includegraphics[width=\textwidth, trim=0.cm 0.9cm 0cm 0cm, clip=True]{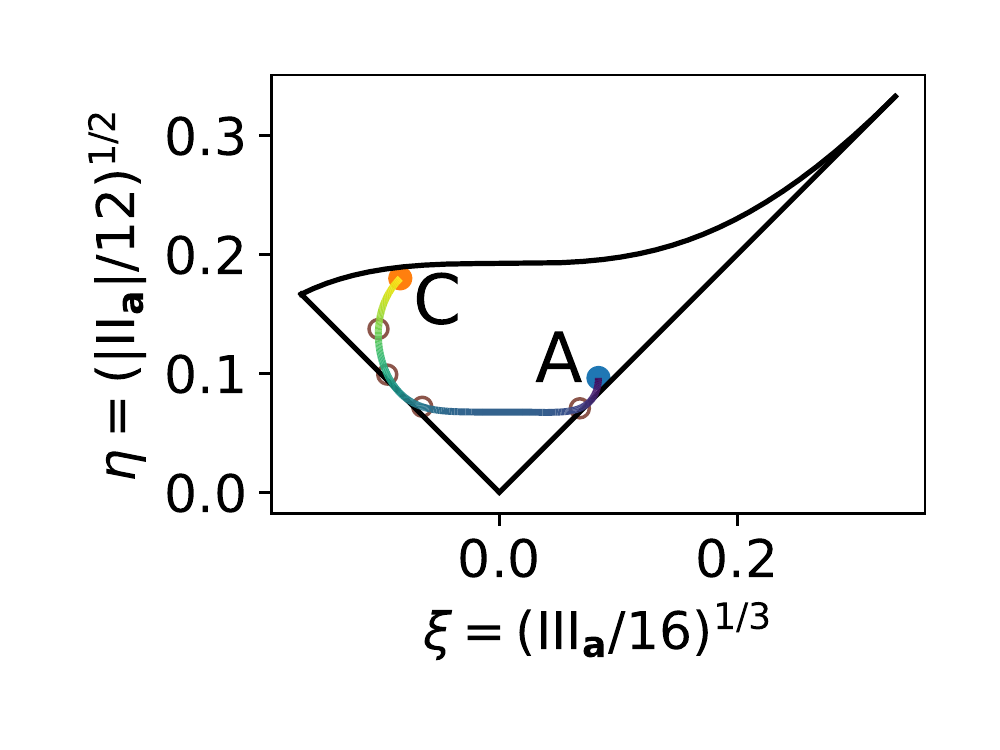}}
             \vspace{-0.6\baselineskip}
         \end{subfloat}
    \end{minipage}
    \begin{subfloat}[\label{fig:tensorEllipsoidAtoB}Reynolds stress tensor ellipsoid]
        {\begin{tikzpicture}
            \node[] at (0,16) {
                \includegraphics[width=0.24\textwidth, trim=3cm 1.8cm 3cm 2cm, clip=True]{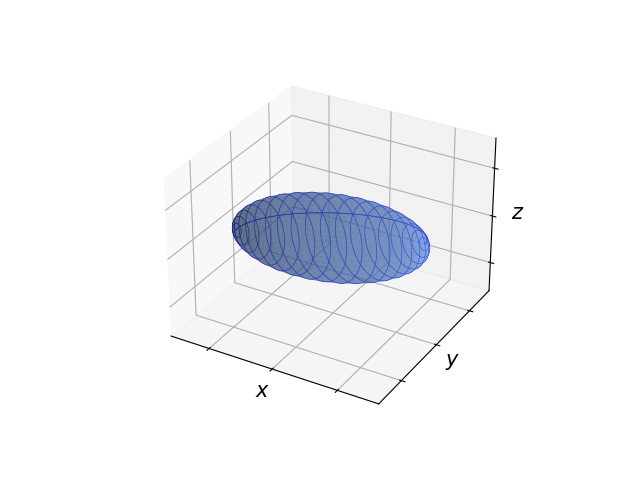}};
            \node[] at (0,12.8) {
                \includegraphics[width=0.24\textwidth, trim=3cm 1.8cm 3cm 2cm, clip=True]{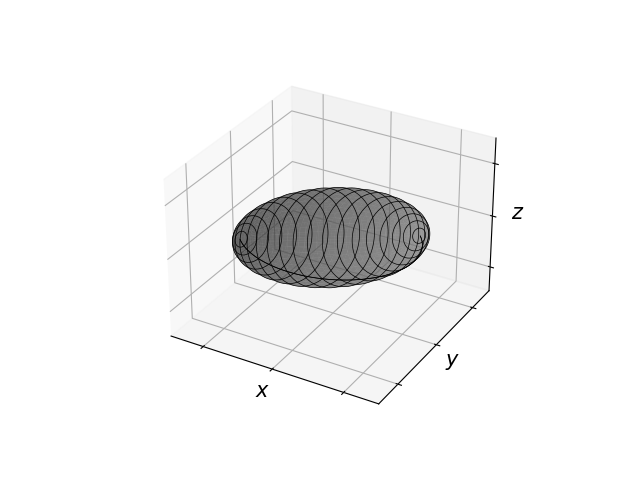}};
            \node[] at (0,9.6) {
                \includegraphics[width=0.24\textwidth, trim=3cm 1.8cm 3cm 2cm, clip=True]{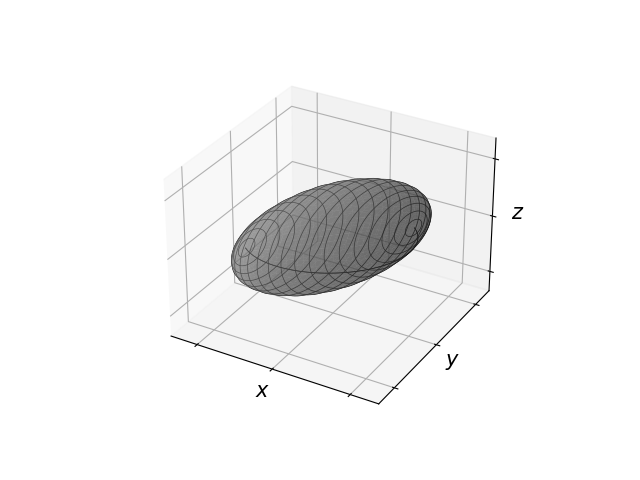}};
            \node[] at (0,6.4) {
                \includegraphics[width=0.24\textwidth, trim=3cm 1.8cm 3cm 2cm, clip=True]{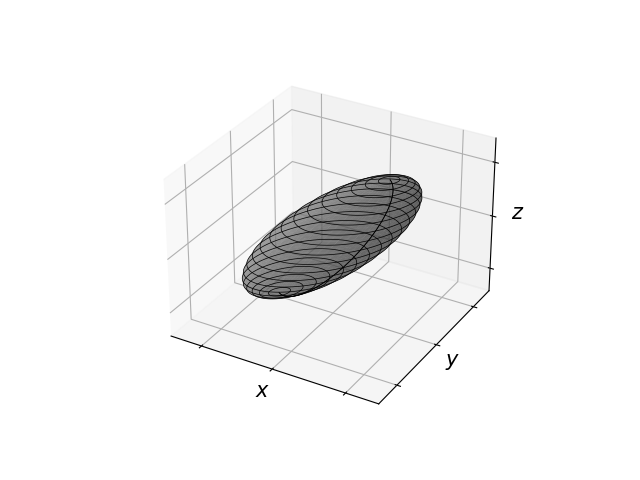}};
            \node[] at (0,3.2) {
                \includegraphics[width=0.24\textwidth, trim=3cm 1.8cm 3cm 2cm, clip=True]{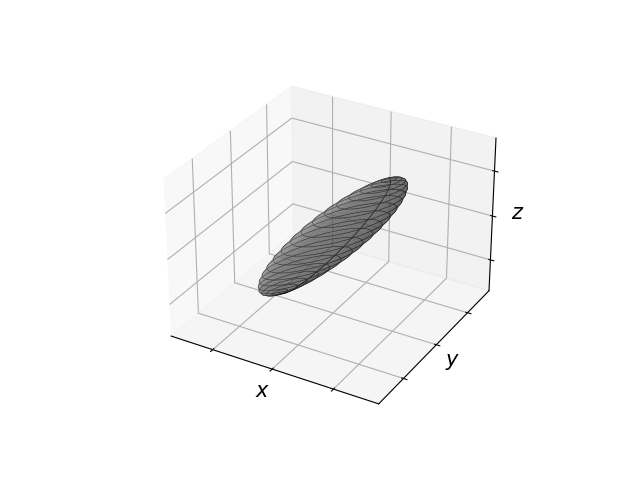}};
            \node[] at (0,0) {
                \includegraphics[width=0.24\textwidth, trim=3cm 1.8cm 3cm 2cm, clip=True]{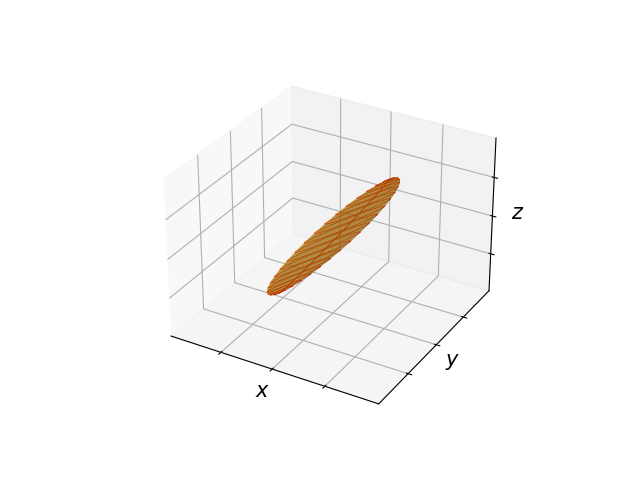}};
            \node[circle, black, draw] at (-1, 17) {$\mathrm{\mathbf{A}}$};
            \node[ black, draw] at (-0.9, 13.6) {\scriptsize $f=0.2$};
            \node[ black, draw] at (-0.9, 10.7) {\scriptsize $f=0.4$};
            \node[ black, draw] at (-0.9, 7.3) {\scriptsize $f=0.6$};
            \node[ black, draw] at (-0.9, 4.0) {\scriptsize $f=0.8$};
            \node[circle, black, draw] at (-1, 0.5) {$\mathrm{\mathbf{C}}$};
        \end{tikzpicture}}  
     \end{subfloat}
     \begin{subfloat}[\label{fig:eigenvectorsAtoB}Eigenvectors of Reynolds stress tensor]
        {\begin{tikzpicture}
            \node[] at (0,16) {
                \includegraphics[width=0.24\textwidth, trim=3cm 1.8cm 3cm 2cm, clip=True]{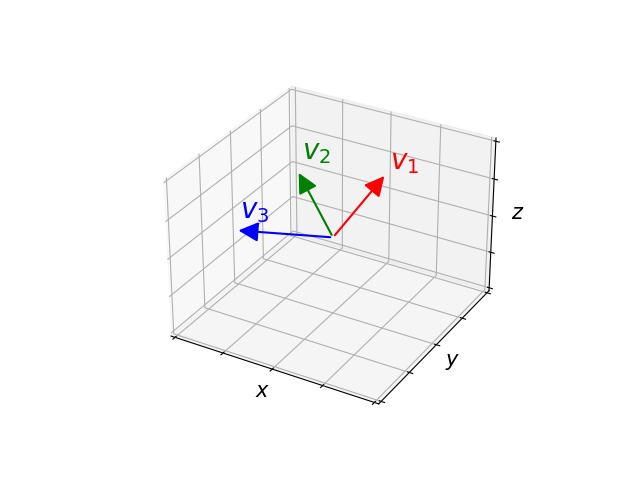}};
            \node[] at (0,12.8) {
                \includegraphics[width=0.24\textwidth, trim=3cm 1.8cm 3cm 2cm, clip=True]{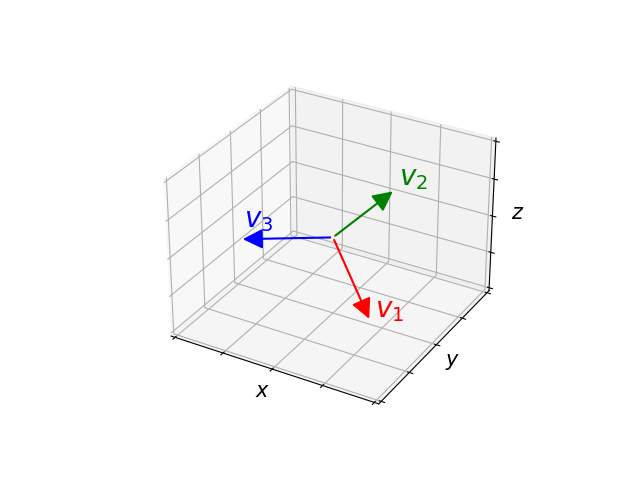}};
            \node[] at (0,9.6) {
                \includegraphics[width=0.24\textwidth, trim=3cm 1.8cm 3cm 2cm, clip=True]{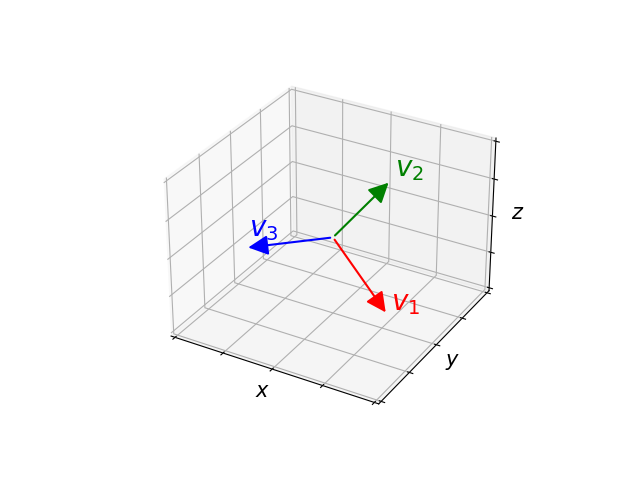}};
            \node[] at (0,6.4) {
                \includegraphics[width=0.24\textwidth, trim=3cm 1.8cm 3cm 2cm, clip=True]{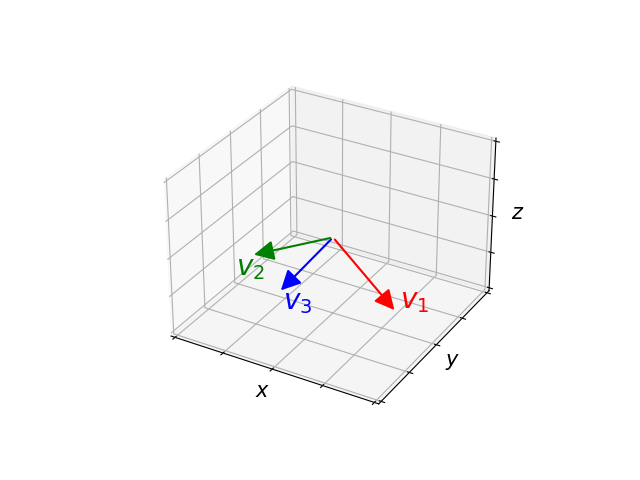}};
            \node[] at (0,3.2) {
                \includegraphics[width=0.24\textwidth, trim=3cm 1.8cm 3cm 2cm, clip=True]{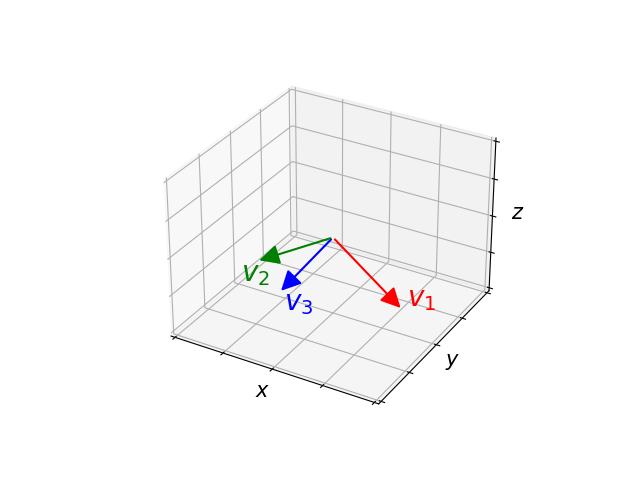}};
            \node[] at (0,0) {
                \includegraphics[width=0.24\textwidth, trim=3cm 1.8cm 3cm 2cm, clip=True]{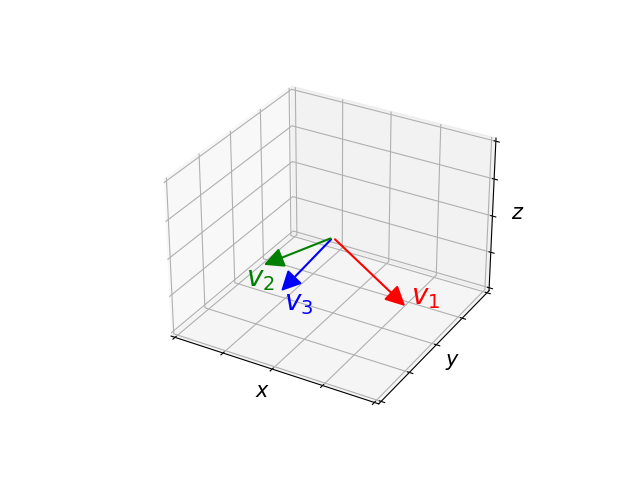}};
            \node[circle, black, draw] at (-1, 17) {$\mathrm{\mathbf{A}}$};
            \node[ black, draw] at (-0.9, 13.6) {\scriptsize $f=0.2$};
            \node[ black, draw] at (-0.9, 10.7) {\scriptsize $f=0.4$};
            \node[ black, draw] at (-0.9, 7.3) {\scriptsize $f=0.6$};
            \node[ black, draw] at (-0.9, 4.0) {\scriptsize $f=0.8$};
            \node[circle, black, draw] at (-1, 0.5) {$\mathrm{\mathbf{C}}$};
        \end{tikzpicture}}
    \end{subfloat}
\caption{Transition from tensor $\mathrm{\mathbf{A}}$ to $\mathrm{\mathbf{C}}$ (defined in \cref{app:tensorABC}) featuring different eigenvector by increasing $f=0...1$ (see  \cref{eq:summationWithFactor}). The intermediate brown-colored states in \protect\subref{fig:baryAtoB}, \protect\subref{fig:invariantsAtoB} and \protect\subref{fig:xiEtaAtoB} correspond to the states with $f\in \text{[}0.2, 0.4, 0.6, 0.8\text{]}$ in \protect\subref{fig:tensorEllipsoidAtoB} and \protect\subref{fig:eigenvectorsAtoB}.}
\label{fig:blendingDifferentEigenvectors}
\end{figure}
Finally, if $f$ is increased incrementally from $0$ to $1$, the resulting states $\mathbf{x}_\mathbf{Z}$ will end up forming a straight line connecting $\mathbf{x}_\mathbf{X}$ and $\mathbf{x}_\mathbf{Y}$, as illustrated in \cref{fig:blendingIdenticalEigenvectors} and especially in \cref{fig:baryAtoC}. For reference, \cref{fig:invariantsAtoC} and \cref{fig:xiEtaAtoC} show the result of linear interpolation in terms of barycentric coordinates in the classical \gls{AIM} and the alternative Anisotropy Invariant Map~\cite{Choi}.

However, the summation of commuting matrices is the\cancelText{ special case} \newText{exception.}\cancelText{for} \newText{Adding up} two arbitrary, positive semi-definite matrices, eigenvector orientation is not preserved and the resulting eigenvalues are not just the sum of the original eigenvalues.
As a consequence, their transformation into barycentric coordinates is not located along the shortest possible path connecting the representation of the anisotropy of the original tensors, as shown in \cref{fig:blendingDifferentEigenvectors}.\cancelText{ By analysing} \newText{Analyzing} the orientation of the \gls{PCS} of each tensor in \cref{fig:tensorEllipsoidAtoB} and \cref{fig:eigenvectorsAtoB}\cancelText{, the transformation of eigenspace is revealed} \newText{reveals the transformation of eigenspace}. The representation in barycentric coordinates shows a perturbation trajectory which connects starting and target point via the sides of the triangle (see \cref{fig:baryAtoB}). Hence, the introduction of a moderation factor violates the original intent of the \gls{EPF} and, in addition to that, affects the plausibility of recent data-driven  machine learning approaches~\cite{Heyse, MathaCF}, relying on the interpolation property with respect to barycentric coordinates.
Moreover, the\cancelText{ limits} \newText{bounds} of the Frobenius inner \newText{matrix} product \cancelText{ based on two matrices} (see \cref{eq:minMaxProduction}) can only be achieved, if the matrices share the same eigenvectors. When applying Reynolds stress eigenvector perturbation in combination with a moderation factor, the resulting turbulent production indeed yields a value within the interval of the inner product defined in \cref{sec:eigenvectorPerturbation}, but does not reach the theoretical limits as the perturbed Reynolds stress tensor features some different eigenvectors compared to the strain-rate tensor.
\\
To sum up,\cancelText{ one of} the concept\cancelText{s} of the \gls{EPF}, which is \newText{perturbing the eigenvalues of the Reynolds stress tensor linearly between the initial state and a} certain limiting state of turbulence, cannot be guaranteed if a moderation factor is introduced as in current implementations. Applying this moderation factor in combination with eigenvector perturbations results in an conceptually unintended state of the anisotropy tensor on the one hand. On the other hand, the intended minimization and maximization of the turbulent production term is no longer guaranteed. 

\subsubsection{\label{sec:proposedApproach} Proposed approach to improve self-consistency}
\begin{figure}
    \begin{subfloat}[\label{channelFlowSetup1}]
         {\includegraphics[scale=0.5]{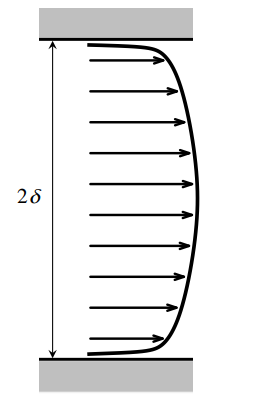}}
     \end{subfloat}
     \begin{subfloat}[\label{channelFlowSetup2}]
         {\includegraphics[scale=0.5]{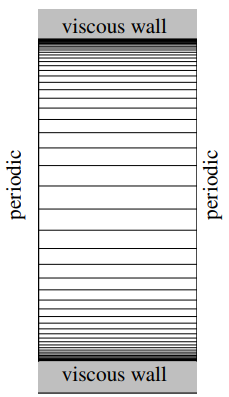}}
     \end{subfloat}
\caption{Turbulent channel flow simulation. (a) Schematic sketch of a fully developed turbulent boundary layer. (b) Mesh (every fourth line shown) and boundary conditions; symmetry is enforced in spanwise direction.}
\label{Bild:channelFlow}
\end{figure}

\begin{figure}
\begin{tikzpicture}
\node[] at (0,0) {\includegraphics[width=0.38\textwidth, trim=0.9cm 0.8cm 0.7cm 0.1cm, clip=True]{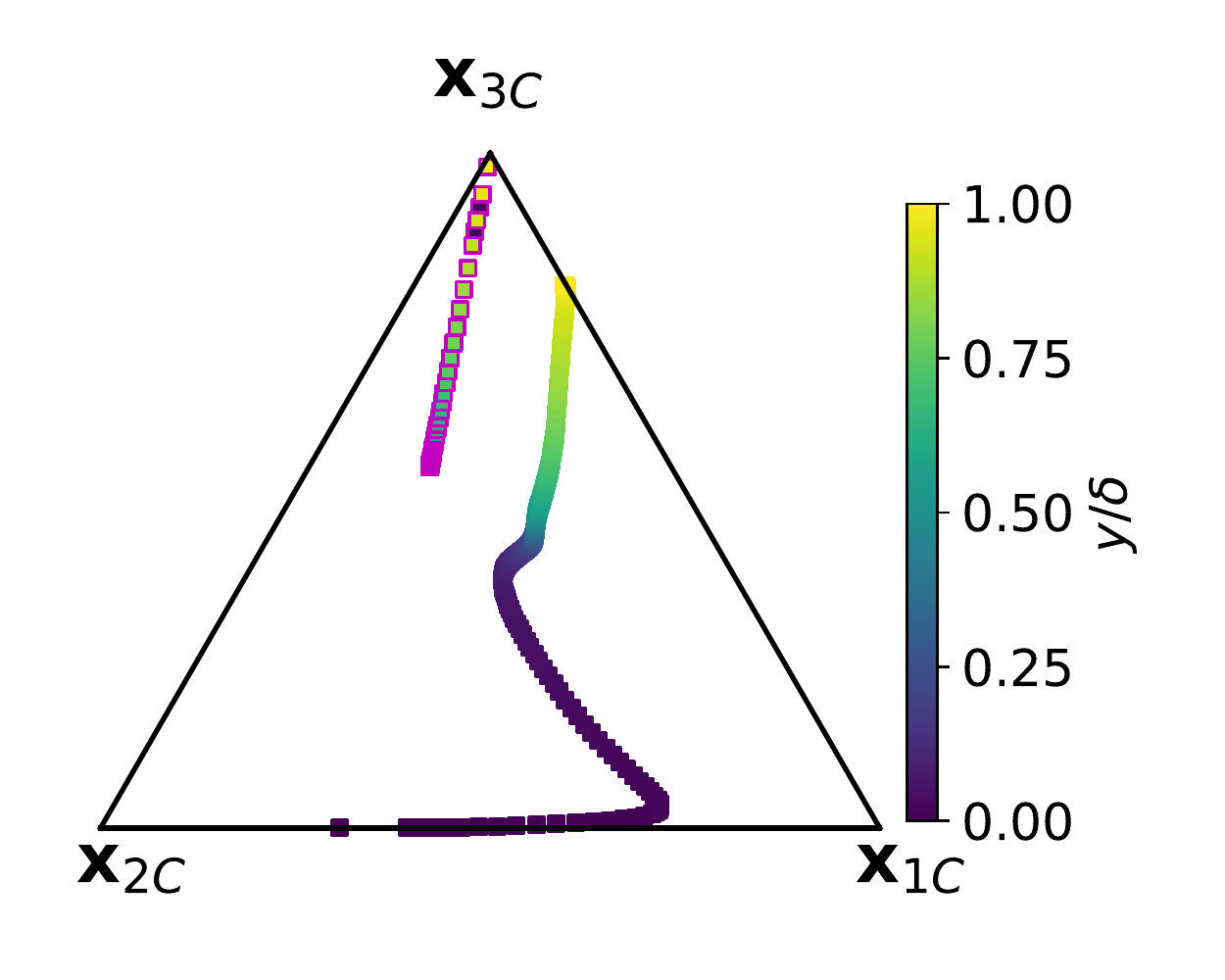}};
\draw [gray!80!white, ->]( -2.3,1.5) -- (-1.1,0.6) node[] {};
\draw [gray!80!white, ->](0.2,1.5) -- ( -0.2,0.1) node[] {};
\node[black] at (-2.5, 1.75) {\footnotesize baseline RANS};
\node[black] at (0.2, 1.75) {\footnotesize DNS};
\end{tikzpicture}
\caption{\label{fig:baryCoordinatesChannelBaselineDNS}Barycentric coordinates of \gls{DNS} data and RANS simulation using baseline turbulence model MenterSST $k$-$\omega$. Data points are colored according to their distance from the wall.}
\end{figure}
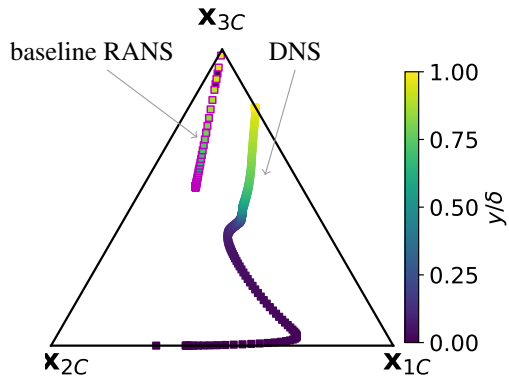

In order to resolve the issues described in \cref{sec:unwantedEffect} the implementation of the \gls{EPF} needs to be changed.
A first step is the removal of the entire idea of applying a moderation factor to adjust the amount of perturbed Reynolds stress tensor according to \cref{eq:moderationFactorReconstruction}. As a consequence, \cancelText{only adjusting} $\Delta_B$ in \cref{eq:perturbationMagnitude}, which controls the amount of perturbation towards the respective limiting state of turbulence, \newText{has to be adjusted,} in order to retain converged \gls{RANS} simulations (see \cref{sec:needForModeration}). This is in contrast to the the common practice of choosing $\Delta_B =1.0$, arguing that there is no physical reason to restrict this value without the usage of data-driven methods or expert knowledge on the flow configuration. In other words, the perturbed Reynolds stress tensor, entering the update of the viscous fluxes and the turbulent production term, in the proposed self-consistent implementation is equal to \cref{eq:spectralDecompositionR*}. Nevertheless, the fundamental idea of the individual perturbation of eigenvalues and eigenvectors, introduced in \cref{sec:eigenvaluePerturbation}, remains the same.
\cancelText{In doing so} \newText{Hereby}, the entire \gls{EPF} in order to quantify the structural uncertainties of turbulence models is formulated in a verified, physics-constrained and self-consistent manner.
\newText{Its implementation in \textit{TRACE} can be subdivided in several steps within each pseudo-time step of steady \gls{RANS}:
\begin{enumerate}
    \item Calculate Reynolds stress tensor based on Boussinesq approximation in \cref{eq:boussinesq}
    \item Determine respective anisotropy tensor (see \cref{eq:decompositionTau}).
    \item Decompose the anisotropy tensor in its eigenvalues and eigenvectors (see \cref{eq:spectralDecompositionAnisotropy}). 
    \item Compute the barycentric coordinates based on eigenvalues of the anisotropy tensor (see \cref{eq:barycentricMapping}). 
    \item Perturb the barycentric coordinates of the anisotropy tensor within physical realizable limits by chosen $\Delta_B$ (see \cref{eq:perturbationMagnitude})
    \item Determine perturbed eigenvalues of the anisotropy tensor with respect to the perturbed barycentric coordinates (see 
    \cref{eq:perturbedEigenvalues}) 
    \item Perturb the eigenvectors of anisotropy/Reynolds stress tensor if turbulent production term should be minimized (see \cref{eq:minMaxProduction}).
    \item Reconstruct the perturbed Reynolds stress tensor according to \cref{eq:spectralDecompositionR*}
    \item Update the viscous fluxes using the reconstructed perturbed Reynolds stress tensor 
    \item Update the turbulence production term using the reconstructed perturbed Reynolds stress tensor  explicitly
\end{enumerate}
}
\newTextTwo{Note: Different types and magnitudes of the perturbations (1C, 2C or 3C; $P_{k_{\mathrm{min}}}$ or $P_{k_{\mathrm{max}}}$; chosen $\Delta_B$ and/or $f$) result in different solutions of the RANS equations from a mathematical point of view regardless of the \gls{EPF} formulation (non-consistent or consistent). However, not every mathematical solution represents a physically meaningful solution (e.g. a solution giving laminar flow in a clearly turbulent domain, or unsteady flow in steady state conditions). Hence, the \gls{EPF} requires certain expert knowledge and engineering practice to determine the appropriate amount of perturbation magnitude ($\Delta_B$ in the consistent formulation) leading to meaningful, converged RANS solutions.}
\newText{
\subsection{\label{sec:channelFlowExample}Application to plane turbulent channel flow}
The uncertainty estimates based on the non-consistent and self-consistent eigenspace perturbation are compared when applied to a canonical turbulent channel flow at $Re_\tau=1000$.
The channel flow is homogeneous in streamwise and spanwise direction. A constant streamwise pressure gradient $\partial p/ \partial x$ is applied to balance the skin friction at the walls. The configuration for simulating this wall-bounded flow is sketched in \cref{Bild:channelFlow}. The mesh has a low-Reynolds resolution ($y^+ \leq 1$) at the solid walls with 100 cells up to the symmetry line in wall-normal direction.
The two-equation Menter SST $k$-$\omega$ turbulence model\cite{Menter}, which belongs to the group of \gls{LEVM}, is considered as the baseline model for the present simulations.
The discrepancies with respect to barycentric coordinates of the \gls{RANS} turbulence model when compared with available  \gls{DNS} data \cite{lee_moser_2015} are moderate in the channel center and start to increase close to the wall due to the strong anisotropy of turbulence (see \cref{fig:baryCoordinatesChannelBaselineDNS}). Due to the fact, that the turbulence model relies on the Boussinesq assumption \cref{eq:boussinesq} and that a velocity gradient in spanwise direction is missing, the Reynolds stress tensor has at least one zero eigenvalue. Hence, the resulting barycentric coordinates are known to be the plane-strain line in the \gls{ABM}.}


\newText{The turbulence model-form uncertainty is quantified applying the \gls{EPF}. In order to demonstrate the implications of using the proposed consistent formulation a relative perturbation strength of $\Delta=0.5$ is used for the consistent formulation, while $\Delta=1.0$ is used for the non-consistent formulation. Consequently, a factor of $f=0.5$ is applied for the non-consistent formulation to moderate the strength for eigenvalues and eigenvector perturbation and to obtain comparable results to the consistent formulation. The streamwise pressure gradient, which was adjusted for the baseline simulation to match the Reynolds number, remains constant throughout the perturbed simulations. This is comparable to the procedure of Emory et al. for a similar test case\cite{Emory}. 
The comparison of the uncertainty estimated by the \gls{EPF} for the streamwise velocity profile of the boundary layer is presented in \cref{Bild:u_profiles}. The simulations featuring eigenvector perturbation are indicated by $P_{k_{\mathrm{min}}}$ (leading to minimized turbulent production), while no eigenvector permutation is applied for $P_{k_{\mathrm{max}}}$ (see \cref{eq:minMaxProduction}). 
\begin{figure}
    \begin{subfloat}[\label{uProfilNonConsistent}]
         {\includegraphics[scale=0.5]{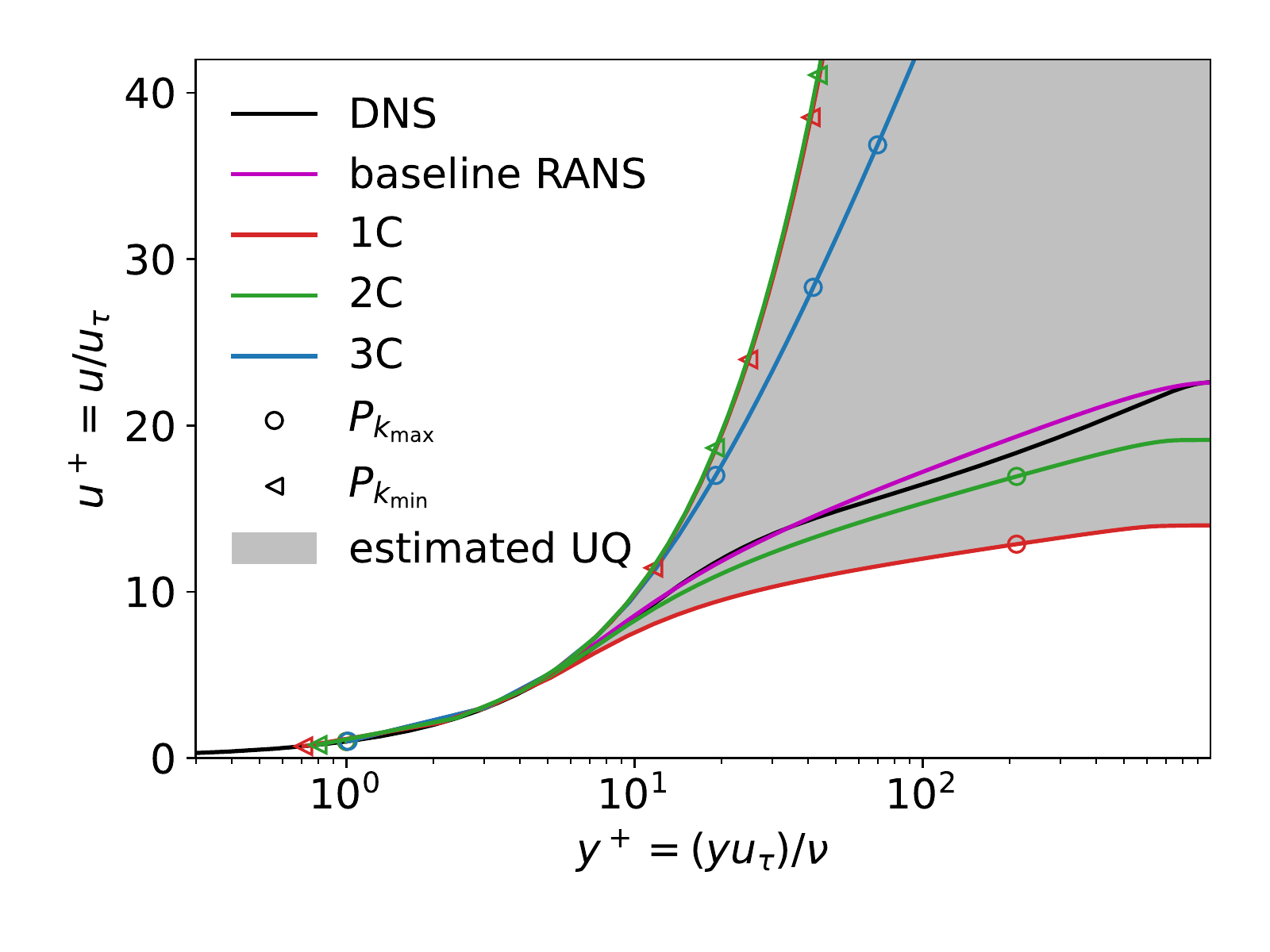}}
     \end{subfloat}
     \begin{subfloat}[\label{uProfilConsistent}]
         {\includegraphics[scale=0.5]{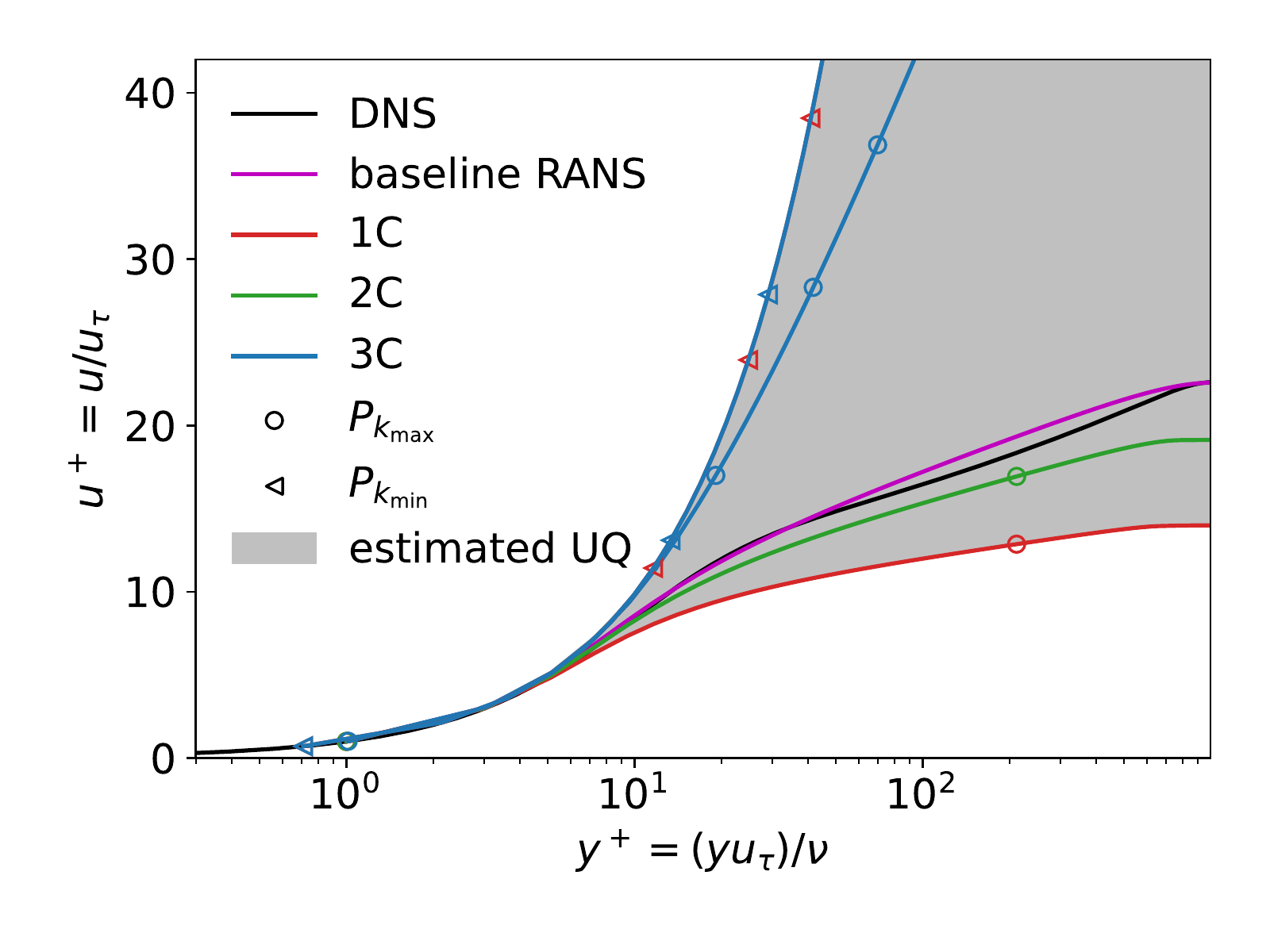}}
     \end{subfloat}
\caption{Comparison of resulting uncertainty bounds for streamwise velocity profile of turbulent channel flow simulation derived by \gls{EPF}. (a) Application of non-consistent formulation of \gls{EPF} using $\Delta_B =1.0$ and $f=0.5$. (b) Application of consistent formulation of \gls{EPF} using $\Delta_B =0.5$.}
\label{Bild:u_profiles}
\end{figure}
Overall, the uncertainty estimate (grey shaded area) of the  boundary layer profile are identical for both formulations. On the one hand, this is because the simulations without any eigenvector perturbation are mathematically equivalent, leading to identical results for \gls{QoI}.
On the other hand, applying eigenvector permutation for the channel flow results in laminarization of the boundary layer. Hence, the laminar velocity profile driven by the selected constant streamwise pressure gradient bounds the uncertainty estimation, regardless of \gls{EPF} formulation or target barycentric coordinate $\mathbf{x}_{(t)}$. Overall, the uncertainty intervals are smaller for previous investigations of the channel flow by Emory et al. \cite{Emory}. To the authors' knowledge and experience, this is because of the fact, that Emory et al. do not explicitly update the turbulent production term based on the perturbed Reynolds stresses. Additionally, as the perturbations for both formulations are uniform throughout the computational domain, it is expected, that by applying an appropriate amount of perturbation strength (e.g. locally varying perturbations with the help of machine learning) the uncertainty estimates would be more adequate.}

\begin{figure}
    \begin{subfloat}[\label{fig:baryNonConsistent}]
         {\includegraphics[width=0.3\textwidth, trim=0.9cm 1cm 0.7cm 1.1cm, clip=True]{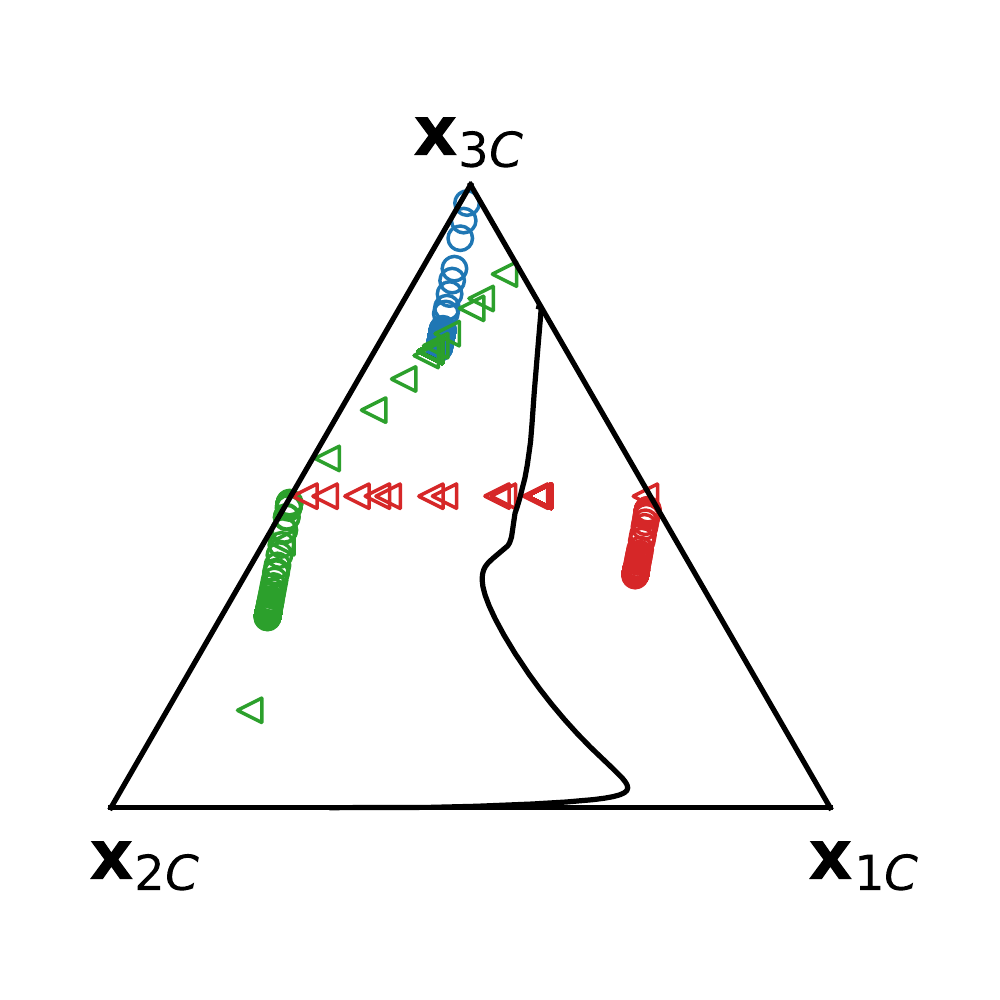}}
     \end{subfloat}
     \begin{tikzpicture}
            \node[] at (0,0) {
                \includegraphics[width=0.085\textwidth, trim=4.2cm 1.7cm 4cm 1cm, clip=True]{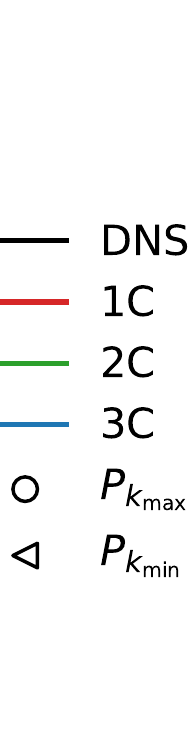}};
    \end{tikzpicture}
     \begin{subfloat}[\label{fig:baryConsistent}]
         {\includegraphics[width=0.3\textwidth, trim=0.9cm 1cm 0.7cm 1.1cm, clip=True]{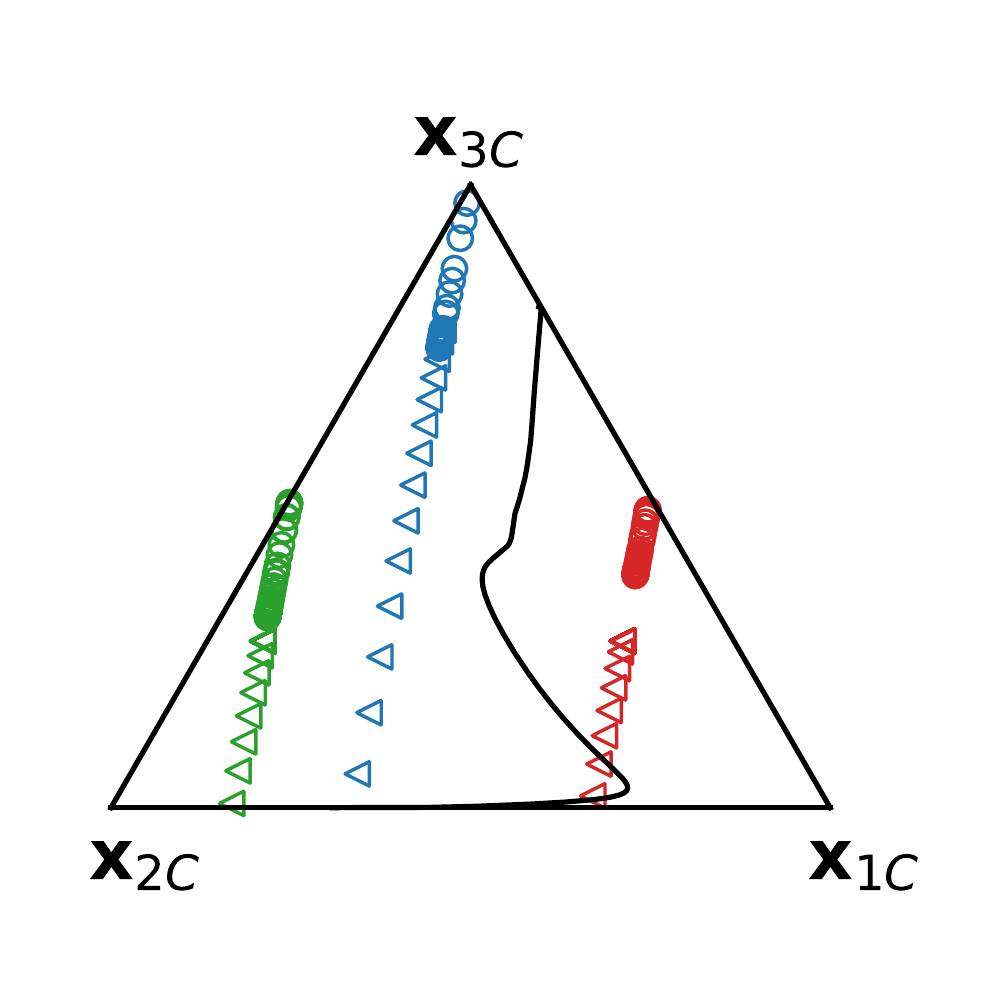}}
     \end{subfloat}
\caption{Comparison of resulting barycentric coordinates of the perturbed Reynolds stress tensors for turbulent boundary layer profiles in \cref{Bild:u_profiles}. (a) Application of non-consistent formulation of \gls{EPF} using $\Delta_B =1.0$ and $f=0.5$. (b) Application of consistent formulation of \gls{EPF} using $\Delta_B =0.5$.}
\label{Bild:baryCoordinatesChannel}
\end{figure}

\newText{In terms of conceptual model verification, the proposed self-consistent formulation guarantees to maintain agreement between the theoretical idea of the \gls{EPF} and the simulation results, which are shown in \cref{Bild:baryCoordinatesChannel}. The final perturbed states of the Reynolds stress tensor anisotropy for simulations aiming at $P_{k_{\mathrm{max}}}$, show the expected, identical perturbed anisotropic states for both \gls{EPF} formulations. \cref{fig:baryNonConsistent} reveals the initial motivation for scrutinizing the consistency of the formulation using a moderation factor in combination with eigenvector perturbation as the RANS solution points for the turbulent boundary layer do not show the intended behaviour for $P_{k_{\mathrm{min}}}$. If a \gls{CFD} practitioner runs a perturbed \gls{RANS} simulation aiming for one of the corners of the barycentric triangle, it is expected that the resulting anisotropic states show respective shifts towards that limiting state of turbulence. The boundary layer solution points of $(\mathrm{1C}, P_{k_{\mathrm{min}}})$ and $(\mathrm{2C}, P_{k_{\mathrm{min}}})$ are located at some unintended states inside the barycentric triangle in \cref{fig:baryNonConsistent} due to summation of two non-commuting tensors. In contrast, the respective simulations using the consistent formulation produces anisotropic states of the Reynolds stress tensor, which are entirely perturbed towards one of the corners of the triangle (keeping in mind, that the unperturbed Reynolds stress tensor is represented by the plane-strain line as in \cref{fig:baryCoordinatesChannelBaselineDNS}).\\
Note: The self-consistent formulation of the \gls{EPF} framework, presented in \cref{sec:proposedApproach}, enables the user to additionally perform the perturbed RANS simulation aiming for $(\mathrm{3C}, P_{k_{\mathrm{min}}})$, which was obsolete in the non-consistent formulation using $\Delta_B=1.0$.}

\section{\label{sec:conclusion}Conclusion \& Outlook}
The \gls{EPF}, that creates perturbed states of the Reynolds stress tensor in order to quantify the structural uncertainties of \gls{RANS} turbulence models, is described in detail, presenting its underlying idea and discussing its practical implementation and usage.
 The present work highlights one shortcoming in the commonly proposed implementation of this framework.
 Due to numerical convergence issues, researchers have suggested to weaken the effect of perturbed Reynolds stress by introducing a moderation factor in previous publications.
 The assessment of the common computational implementation reveals, that the basic concept of the \gls{EPF} is not correctly represented in that case. The introduction of a separate moderation factor may cause unintended behaviour and violate \gls{EPF}'s self-consistency\cancelText{ cannot be guaranteed anymore}. Therefore, the present paper presents a self-consistent way of formulating the Reynolds stress tensor perturbation framework, as the significance of reasonable physics-constrained Uncertainty Quantification methodologies is indisputable.
This formulation has recently been implemented in \gls{DLR}'s \gls{CFD} solver suite \textit{TRACE}. \newText{The analysis of the results based on the proposed eigenspace perturbation formulation when applied to turbulent channel flow verifies its benefits with respect to the interpretability of the uncertainty estimates. In the near future the framework}\cancelText{ and} will be applied to quantify the uncertainties for \newText{more complex flows}\cancelText{ turbomachinery flows in the near future} \newText{for which the differences between the non-consistent and the self-consistent formulation are expected to be greater for \gls{QoI}}. Moreover, ongoing research focusing on determining the Reynolds stress tensor perturbation by the use of data-driven machine learning practises will benefit from verified self-consistent implementation of the framework as well.

\begin{acknowledgments}
The project on which this paper is based was funded by the German Federal Ministry for Economic Affairs and Climate Action under the funding code 03EE5041A.
The authors are responsible for the content of this  publication.
\end{acknowledgments}

\section*{Data Availability Statement}

The data that support the findings of this study are available
from the corresponding author upon reasonable request.

\appendix

\section{\label{app:eigenvectors} Properties of the sum of two tensors featuring identical eigenvectors in terms of eigenspace}

Let $\phi_i$ be the eigenvalues of tensor $\mathbf{X}$ and $\psi_i$ be the eigenvalues of tensor $\mathbf{Y}$. Both tensors share the same eigenvectors $\mathrm{w}_i$.
Therefore, we know, that the relationships
\begin{equation}
    \label{eq:proof1Eq1}
    \begin{split}
       \mathbf{X} \mathbf{w}_{i} = \phi_i \mathrm{w}_i \quad i ={1,2,3} \\
       \mathbf{Y} \mathbf{w}_{i} = \psi_i \mathrm{w}_i \quad i ={1,2,3} 
    \end{split}
\end{equation}
are satisfied.
The summation of $\mathbf{X}$ and $\mathbf{Y}$ leads to:
\begin{equation}
    \label{eq:proof1Eq2}
    \begin{split}
       \left(\mathbf{X}+\mathbf{Y}\right)\mathbf{w}_{i} &= \mathbf{X}\mathbf{w}_{i} + \mathbf{Y}\mathbf{w}_{i} \quad i ={1,2,3}\\
        &= \phi_i \mathbf{w}_i + \psi_i \mathbf{w}_i  \ \text{(cf. \cref{eq:proof1Eq1})} \\
        &= \left(\phi_i+ \psi_i\right) \mathbf{w}_i
    \end{split}
\end{equation}
Consequently, the resulting sum features identical eigenvectors as well and its eigenvalues are the sum of the individual eigenvalues. 

\section{\label{app:positiveSemiDefinite} Transferability of definiteness related to the sum of two positive semi-definite tensors}

Tensor $\mathbf{X}$ and tensor $\mathbf{Y}$ are positive semi-definite, which means
\begin{equation}
       \forall \mathbf{u} \in \mathbb{R}^n, \ \mathbf{u}^T \mathbf{X} \mathbf{u} \geq 0, \ \mathbf{u}^T \mathbf{Y} \mathbf{u} \geq 0 \ \text{.}
\end{equation}
The sum of $\mathbf{X}$ and $\mathbf{Y}$ can be distributed based on the laws of tensor multiplication
\begin{equation}
    \label{eq:proofSumSemiDefinit}
       \forall \mathbf{u} \in \mathbb{R}^n,  \mathbf{u}^T \left(\mathbf{X}+\mathbf{Y}\right) \mathbf{u} = \mathbf{u}^T \mathbf{X}\mathbf{u} + \mathbf{u}^T \mathbf{Y} \mathbf{u} \geq 0
\end{equation}
Consequently, the sum of two positive semi-definite tensors is positive semi-definite as well.

\section{\label{app:interpolationProperties} Interpolation properties of two scaled tensor with respect its location in barycentric coordinates}

Let $\phi_1\geq\phi_2 \geq\phi_3$ be the eigenvalues of the anisotropic part of the (3,3)-tensor $\mathbf{X}$ and $\psi_1\geq\psi_2 \geq\psi_3$ be the eigenvalues of the anisotropic part of the (3,3)-tensor $\mathbf{Y}$.
The eigenvalues of the summation of the scaled tensors 
\begin{equation}
        \mathbf{Z}  = \left(1-f\right)\mathbf{X} + f \mathbf{Y}
\end{equation}
are $\sigma_i = \left(1-f\right)\phi_i+f\psi_i$. The barycentric coordinates are
\begin{align}
        \mathbf{x}_\mathbf{Z} =& \mathbf{x}_{\mathrm{1C}}\frac{1}{2}\left[\sigma_1-\sigma_2\right]+ \mathbf{x}_{\mathrm{2C}}\left[\sigma_2-\sigma_3\right]+ \mathbf{x}_{\mathrm{3C}} \left[\frac{3}{2}  \sigma_3+1\right] 
        \\
        \begin{split}
        = & \mathbf{x}_{\mathrm{1C}}\frac{1}{2}\left[\left(\left(1-f\right)\phi_1+f\psi_1\right)-\left(\left(1-f\right)\phi_2+f\psi_2\right)\right]\\
        &+ \mathbf{x}_{\mathrm{2C}}\left[\left(\left(1-f\right)\phi_2+f\psi_2\right)-\left(\left(1-f\right)\phi_3+f\psi_3\right)\right]\\
        &+ \mathbf{x}_{\mathrm{3C}} \left[\frac{3}{2}  \left(\left(1-f\right)\phi_3+f\psi_3\right)+1\right]
        \end{split}
        \\
        \begin{split}
        = & \mathbf{x}_{\mathrm{1C}}\frac{1}{2}\left[\left(\left(1-f\right)\phi_1+f\psi_1\right)-\left(\left(1-f\right)\phi_2+f\psi_2\right)\right]\\
        &+ \mathbf{x}_{\mathrm{2C}}\left[\left(\left(1-f\right)\phi_2+f\psi_2\right)-\left(\left(1-f\right)\phi_3+f\psi_3\right)\right]\\
        &+ \mathbf{x}_{\mathrm{3C}} \left[\frac{3}{2}  \left(\left(1-f\right)\phi_3+f\psi_3\right)+1\newText{-f+f}\right]\
        \end{split}
        \\
        \begin{split}
        = & \mathbf{x}_{\mathrm{1C}}\frac{1}{2}\left[\left(1-f\right)\left(\phi_1-\phi_2\right)\right] \\
        &+ \mathbf{x}_{\mathrm{2C}}\left[\left(1-f\right)\left(\phi_2-\phi_3\right)\right] \\
        &+ \mathbf{x}_{\mathrm{3C}} \left[\frac{3}{2} \left(1-f\right)\left(\phi_3+1\right)\right]\\
        &+ \mathbf{x}_{\mathrm{1C}}\frac{1}{2}\left[f\left(\psi_1-\psi_2\right)\right] \\
        &+ \mathbf{x}_{\mathrm{2C}}\left[f\left(\psi_2-\psi_3\right)\right] \\
        &+ \mathbf{x}_{\mathrm{3C}} \left[\frac{3}{2} f\left(\psi_3+1\right)\right]
    \end{split}
    \\
    = & \left(1-f\right) \mathbf{x}_\mathbf{X} + f \mathbf{x}_\mathbf{Y}
\end{align}
Consequently, the projection onto barycentric coordinates preserves the ability to interpolate linearly between two initial states in the \gls{ABM}.

\section{\label{app:tensorABC} Example tensors used in this paper}

The positive semi-definite tensor $\mathbf{A}$ is defined as
\begin{equation}
        \mathbf{A} = 
        \begin{pmatrix}
         2 & 0.5 & -0.5\\
         0.5 & 2.5 & -0.5 \\
         -0.5 & -0.5 & 1.5 \\
        \end{pmatrix}  \ \text{,}
\end{equation}
with a set of eigenvalues $\rho_{i_\mathbf{A}}$ and eigenvectors $\mathbf{v}_{i_\mathbf{A}}$.\\
Tensor $\mathbf{C}$, which is positive semi-definite as well, reads
\begin{equation}
        \mathbf{C} = 
        \begin{pmatrix}
         1 & 0.5 & 1.5\\
         0.5 & 2 & 0 \\
         1.5 & 0 & 3 \\
        \end{pmatrix} \ \text{.}
\end{equation}
The respective set of eigenvalues is $\rho_{i_\mathbf{C}}$ and eigenvectors are $\mathbf{v}_{i_\mathbf{C}}$.\\
Tensor $\mathbf{B}$ is constructed using the \gls{PCS} defined by the eigenvectors of $\mathbf{A}$ and the eigenvalues of $\mathbf{C}$ 
\begin{equation}
\begin{split}
        \mathbf{B} = v_{in_\mathbf{A}} 
        \begin{pmatrix}
         \rho_{1_\mathbf{C}} & 0 & 0\\
         0 & \rho_{2_\mathbf{C}} & 0 \\
         0 & 0 & \rho_{3_\mathbf{C}} \\
        \end{pmatrix} 
        v_{jl_\mathbf{A}}
        \approx 
        \begin{pmatrix}
         2.19 & 0.55 & -1.11\\
         0.55 & 3.02 & -0.83 \\
         -1.11 & -0.83 & 0.79 \\
        \end{pmatrix} 
\end{split}
\end{equation}

\bibliography{aipsamp}
\end{document}